\documentclass[final,5p,times,twocolumn,authoryear]{elsarticle}

\usepackage[pagewise]{lineno}

\usepackage{amssymb}
\usepackage{amsmath}
\usepackage{longtable}
\usepackage{multirow}
\usepackage{array}
\usepackage{siunitx}
\usepackage{xcolor}
\usepackage{textgreek}
\usepackage[utf8]{inputenc}
\usepackage[english]{babel}

\usepackage{hyperref}
\hypersetup{
    unicode, 
    colorlinks=true,
    linkcolor=linkcolor,
    citecolor=linkcolor,
    filecolor=linkcolor,
    urlcolor=linkcolor,
}
\usepackage{color,colortbl}
\definecolor{linkcolor}{rgb}{0.0,0.3,0.5}
\usepackage{tensind}
\DeclareGraphicsExtensions{.bmp,.png,.jpg,.pdf}
\usepackage{verbatim}
\usepackage[normalem]{ulem}
\usepackage{orcidlink}
\usepackage{soul}
\usepackage{pdflscape}
\usepackage{multicol}
\usepackage{caption}
\def\crdb{{\tt CRDB}}

\newcommand{\nubase}{{\tt NUBASE~2020}}
\newcommand{\usine}{{\tt USINE}}

\newcommand{\Ga}{\ensuremath{\Gamma^{\rm att}}}
\newcommand{\Gb}{\ensuremath{\Gamma^{\beta}}}
\newcommand{\GEC}{\ensuremath{\Gamma^{\rm EC}}}
\newcommand{\GECtilde}{\ensuremath{\tilde{\Gamma}^{\rm EC}}}
\newcommand{\Gi}{\ensuremath{\Gamma^{\rm inel}}}
\newcommand{\Gp}{\ensuremath{\Gamma^{\rm prod}}}
\newcommand{\Gs}{\ensuremath{\Gamma^{\rm str}}}

\newcommand{\Ntwolev}{\ensuremath{N_{2~{\rm lev}}}}
\newcommand{\Nonelev}{\ensuremath{N_{1~{\rm lev}}}}

\newcommand{\Ftwolev}{\ensuremath{\psi_{2~{\rm lev}}}}
\newcommand{\Fonelev}{\ensuremath{\psi_{1{\rm lev}}}}
\newcommand{\siga}{\ensuremath{\sigma^{\rm att}}}
\newcommand{\sigs}{\ensuremath{\sigma^{\rm str}}}
\newcommand{\Ta}{\ensuremath{t_{\rm att}}}
\newcommand{\Td}{\ensuremath{t_{\rm diff}}}
\newcommand{\Ti}{\ensuremath{t_{\rm inel}}}

\newcommand{\Ts}{\ensuremath{t_{\rm str}}}
\newcommand{\TEC}{\ensuremath{t_{\rm EC}}}

\journal{Astroparticle Physics}

\begin{document}

\begin{frontmatter}

\title{Revisiting electron-capture decay for Galactic cosmic-ray data}
\author[label1]{M.~Borchiellini\orcidlink{0009-0006-3805-2983}}
\ead{m.borchiellini@rug.nl}
\affiliation[label1]{
  organization={Kapteyn Astronomical Institute, University of Groningen}, 
  addressline={Landleven 12}, 
  city={Groningen},
  postcode={9747 AD}, 
  country={The Netherlands}
}
\author[label2]{D.~Maurin\orcidlink{0000-0002-5331-0606}}
\ead{dmaurin@lpsc.in2p3.fr}
\affiliation[label2]{
  organization={LPSC, Université Grenoble Alpes, CNRS/IN2P3},
  addressline={53 avenue des Martyrs}, 
  city={Grenoble},
  postcode={38026}, 
  country={France}
}

\author[label1]{M.~Vecchi\orcidlink{0000-0002-5338-6029}}
\ead{m.vecchi@rug.nl}

\begin{abstract}
  Electron-capture (EC) unstable species in Galactic cosmic rays constrain the time elapsed between nucleosynthesis and acceleration. They have also been advocated as tracers of reacceleration or gas inhomogeneities during their transport. The number of EC-unstable species grows with mass, with an expected EC-decay impact more important for larger atomic number and lower energy.
  We revisit the modelling of EC decay and its detectability in the context of recent unmodulated low-energy (Voyager) and high-precision data for heavy (AMS-02) and very-heavy nuclei (ACE-CRIS, CALET and Super-TIGER). 
  We solve the transport equation for a multi-level configuration (up to any number of electrons attached) in the diffusion and leaky-box models.  
  Their decayed fractions are found to be qualitatively similar but with very different absolute fluxes.
  We check that the standard two-level approximation, wherein the cosmic-ray nucleus is fully ionised or with one electron attached, is sufficient for most situations.
  We find that the impact of EC-decay is negligible in current data, except possibly for fluxes or ratios involving $^{51}$Cr, $^{55}$Fe, and Co. 
  These conclusions are robust against significant uncertainties in the attachment and stripping cross-sections.
  This first analysis calls for further investigation, as several forthcoming projects (e.g., TIGERISS) are targeting $Z>30$ cosmic rays. 
\end{abstract}



\end{frontmatter}


\section{Introduction}
\label{intro}
Most cosmic-ray (CR) data and studies focus on stable species, whose abundances and spectra are used to unveil their origin, acceleration, and transport~\citep[e.g.,][]{2007ARNPS..57..285S, 2018ARNPS..68..377T}. Unstable isotopes in CRs are rarer, but bring complementary information. The best known radioactive CRs are spontaneous $\beta$-unstable species with half-life at rest $t_{1/2}$ in the Myr range (and half-life in flight $\gamma t_{1/2}$, with $\gamma$ the Lorentz factor). These species are commonly referred to as CR clocks because, together with their associated ratios (e.g., $^{10}$Be and the $^{10}$Be/$^9$Be ratio), they allow constraining the confinement time, or alternatively, the halo size of the Galaxy~\citep{2002A&A...381..539D, 2020A&A...639A..74W, 2020PhRvD.101b3013E, 2022A&A...667A..25M, 2024PhRvD.109h3036Z}. They are also sensitive to the presence of spatial inhomogeneities during their transport or in the surrounding gas distribution~\citep{1997AdSpR..19..787P, 1998A&A...337..859P, 2001AdSpR..27..743S, 2002A&A...381..539D, 2023MNRAS.526..160J, 2025JCAP...02..062D}.

Unlike spontaneous decay, electron-capture (EC) decay requires an electron from the innermost shell to interact within the nucleus to convert a proton into a neutron. EC decay in CRs is thus a multistep process with competing physical processes at play. Indeed, an electron must first be attached to the CR and then EC-decay must occur before the electron is stripped from the ion.
Above GeV/n energies, electron stripping becomes more efficient than attachment: EC decay is effectively suppressed and CRs are fully ionised.
This property can be used to estimate the time $t_{\rm elapsed}$ between nucleosynthesis and acceleration \citep{1973ICRC....1..546C, 1975ApJ...200L..75C, 1973ICRC....1..534R}. Freshly synthesised CRs at rest have available electrons and can undergo EC decay, whereas they are fully stripped and cannot decay via EC after the acceleration stage: consequently, the fraction of nuclei that have decayed via EC reflects the time they spent at low energy before acceleration. For instance, the EC decay of $^{57}$Co ($t_{1/2}=272$~days) measured in the associated elemental fluxes allowed to set $t_{\rm elapsed}\gtrsim {\cal O}(\rm yr)$ \citep{1978ApJ...219..753S, 1979ApJ...228..582T, 1985ApJ...294..441D, 1990ApJ...348..608W}, while the EC decay of $^{59}$Ni ($t_{1/2}=81$\,kyr) in isotopic ratio data allowed to reach the lower limit $t_{\rm elapsed}\gtrsim 10^4-10^5$\,yrs~\citep{1992ApJ...390L..99L, 1993ApJ...405..567L, 1997AdSpR..19..747L, 1999ApJ...523L..61W}. This lower bound was recently nicely complemented by an upper bound $t_{\rm elapsed}\lesssim 2.6$\,Myr, from the observations of a few $\beta$-unstable $^{60}$Fe ($t_{1/2}=2.62$\,Myr) with the ACE-CRIS instrument~\citep{2016Sci...352..677B}.

EC-unstable species can also be used to trace energy changes during their journey, such as reacceleration in the Galaxy or energy losses during Solar modulation~\citep{1973ICRC....1..534R}. The idea here is to take advantage of the energy dependence of the attachment cross-section (the smaller the energy, the larger the cross-section). As a result, the probability of measuring a smaller (resp. larger) flux than expected, is linked to the energy gained (resp. lost) during the CR journey.
For instance, the interpretation of measured ratios of daughter/parent EC-decay species ($^{49}$Ti/$^{49}$V and $^{51}$V/$^{51}$Cr) initially concluded on the presence of a $100$\,MeV/n reacceleration~\citep{1998A&A...336L..61S, 1999ICRC....3...33C, 2001SSRv...99...41C}. This evidence was later disfavoured when accounting for the production cross-section uncertainties of these species~\citep{2001AdSpR..27..737J, 2003ApJS..144..153W, 2005PhDT.........6S}.
These ratios are measured at the top-of-atmosphere (TOA), corresponding to interstellar (IS) energies $E_{\rm k/n}^{\rm IS}\approx E_{\rm k/n}^{\rm TOA}+\Delta E_{\rm k/n}$, with $\Delta E_{\rm k/n}$ an average loss in the Solar cavity \citep{2004JGRA..109.1101C}. With ACE-CRIS data taken at both solar minimum ($\Delta E_{\rm k/n}\sim 200$\,MeV/n) and maximum ($\Delta E_{\rm k/n}\sim350$\,MeV/n), \citet{2003JGRA..108.8033N} argued for the detection of the energy-dependent EC decay. The same data were also used to set indirect constraints on the CR low-energy IS flux shape~\citep{ 2004AIPC..719..127M, 2007ApJ...663.1335C}, before direct measurements outside the Solar cavity became available~\citep{2013Sci...341..150S, 2019NatAs...3.1013S, 2016ApJ...831...18C}.

EC-decay species were also invoked to trace possible inhomogeneities in the IS medium (ISM) where CR travel, as ions cannot pick up electrons in depleted zones~\citep{1985Ap&SS.114..365L}, therefore suppressing EC decay. This could trace for instance the time of our last passage in the spiral arms~\citep{2017ApJ...851..109B}, or the ionic state of nuclei could be sensitive to the very local inhomogeneities \citep{1995ApJ...438L..83T}. There is also an interesting link with $\gamma$-ray spectroscopy~\citep{2006NuPhA.777...70D}: pure EC-decay species, whose daughter is a $\gamma$-ray emitter, enable to inspect the deduced Galactic CR (GCR) source abundance and direct production of $^{44}$Ti~\citep{2018ApJ...863...86B}.
However, beside the last two studies, almost no work was published on EC-decay species in GCRs in the last 20~years. Interested readers are referred to the reviews of that time for more details~\citep{1985Ap&SS.114..365L,2001SSRv...99...27M}.

The interest in EC-unstable CR species should be renewed soon, though. Indeed, a wealth of experiments are planned to explore, above a few hundreds of MeV/n, species with atomic number $Z>30$, i.e., in a region with numerous EC-unstable species. Actually, some of Super-TIGER data are still being analysed \citep{2021cosp...43E1335S, 2023HEAD...2030305R}, while the TIGER--ISS is scheduled for launch in 2027 \citep{2024icrc.confE.171R}, and the NUCLEON-2~\citep{2019AdSpR..64.2610B, 2021PPNL...18..217V}
and HERO~\citep{2019AdSpR..64.2619K, 2023PPNL...20..637K, 2024PAN....87..151P} satellite missions are envisioned in the next decade (see, for instance Fig.~8 of \citealt{2025arXiv250316173M}).
Besides, ongoing experiments provided measurements with unprecedented precision, in regimes where the impact of EC decay could be directly detected, including, at GeV/n TOA energies, elemental fluxes measured up to Fe at the few percent-level precision by AMS-02~\citep{2021PhR...894....1A}, elemental ratios $26\leq Z\leq40$ measured by Super-TIGER~\citep{2016ApJ...831..148M} and $14\leq Z\leq44$ by CALET
\citep{2025ApJ...988..148A},
isotopic ratios $29<Z<38$ by ACE-CRIS at a few hundred of MeV/n ~\citep{2022ApJ...936...13B}, and last but not least, IS fluxes up to Ni by Voyager down to a few tens of MeV/n~\citep{2016ApJ...831...18C}.
To take a broader view, we recall that EC decay during propagation impacts the derived GCR source abundances \citep{1985Ap&SS.114..365L}, and that the latter are key to connect to the various nucleosynthesis production processes leading to ultra-heavy nuclei~\citep{2003PhR...384....1A, 2007PhR...450...97A, 2017ARNPS..67..253T, 2023A&ARv..31....1A}.

On the modelling side, EC decay has not been discussed much either, recently, with the data interpreted in the leaky-box or weighted-slab approximation. To track EC-decay species, these models relied on a 2-level description of the transport equation, with a fully stripped and a hydrogen-like ion state only~\citep{1984ApJ...279..144L}. The same 2-level description is also used in the diffusion model GALPROP for $Z\leq30$ species~\citep{2022ApJS..262...30P}. In contrast, EC decay is not included in the DRAGON code~\citep{2018JCAP...07..006E}, the authors arguing that EC decay has no impact on the calculated fluxes (which they checked by running the GALPROP code).

For all these reasons, it is important to revisit the modelling and impact of EC decay for GCR fluxes, especially in the context of recent $Z\leq40$ and forthcoming $Z>30$ high-precision data. In Sect.~\ref{sec:Transport-Timescales}, we recall the GCR transport equation and the specifics of EC decay, highlighting the various timescales involved. In Sect.~\ref{sec:Results}, we present our results concerning the generic impact of EC decay and the validity of the two-level ion approximation. In Sect.~\ref{sec:sensibility_data}, we discuss the detectability prospects of EC decay in current GCR isotopic and elemental data. In Sect.~\ref{sec:Conclusions}, we conclude and draw the next steps of this study.
For the sake of readability, we postponed the most technical details in Appendices:  \ref{app:sig_att_str} recaps the formulae for single electron attachment and stripping in the K-shell and their generalisation; \ref{app:solutions} provide analytical solutions for the differential density of EC-unstable species considering any level of attached electrons, for both the leaky-box model (hereafter, LBM) and 1D diffusion model (hereafter, 1D-DM).

\section{Transport equation and timescales}
\label{sec:Transport-Timescales}
GCR transport is described by a  diffusion-advection equation, which includes spatial transport, momentum transport, and various source and sink terms. 

\subsection{From differential density to modulated fluxes (without EC)}
The differential density $N_i$ of a stable or $\beta$-unstable species $i$ as a function of energy $E$ is given by~\citep[e.g.,][]{1990acr..book.....B}
\begin{equation} 
\label{eq:transport_gen}
\begin{split}
    - \vec{\nabla} \left[ D(E) \vec{\nabla} N_i - \vec{V}_c N_i\right] +  \frac{\partial}{\partial E} \left[ b_{\rm tot}(E)N_i - (v/c)^2 K_{pp} \frac{\partial N_i}{\partial E}\right] \\
    + \left(\Gi_i +\Gb_i\right) N_i
     = q_i + \sum_{j>i}\Gamma^{\rm prod}_{j\to i} N_j+ \sum_{k>i}\Gb_{k\to i} N_k \,.
\end{split}
\end{equation}
In the first line, from left to right, the transport terms are: spatial diffusion $D$ (CR scattering off the magnetic turbulence), convective transport $V_c$, energy losses (ionisation, Coulomb, and adiabatic) and gains (momentum diffusion $K_{pp}$). In the second line, the CR destruction (left-hand side) and production from heavier species (right-hand side) involve, respectively, the inelastic cross-section of CR species $i$ on the ISM targets $t$, and the spallation cross-section of $i$ on $t$ to produce $j$:
\begin{equation}
    \label{eq:gamma_nuc}
    \displaystyle \Gi_i= \sum_{t\,\in\, \rm ISM}\, n_t \, v_i \,\sigma_{i+t}^{\rm inel}
    \quad {\rm and} \quad
    \displaystyle \Gp_{j\to i}= \sum_{t\,\in\, \rm ISM}\, n_t \, v_i \,\sigma_{j+t\to i}^{\rm prod}\,.
\end{equation}
It also involves $\beta$ decay (half-life $t_{1/2}$) as a sink (decay of $i$) or a source (decay of CR $k$ into $i$),
\begin{equation}
  \label{eq:gamma_decay}
  \Gb = \frac{\ln(2)}{\gamma t_{1/2}}\,.
\end{equation}
We set $n_{\rm ISM} = \sum_t n_t = 1\,\text{cm}^{-3}$ considering a contribution to the ISM number density from H and He of 90\% and 10\%, respectively~\citep[e.g.,][]{2020CoPhC.24706942M}.
Finally, the injection rate $q_i$ describes the primary production.

This transport equation can be solved by different approaches and associated codes: numerically after discretisation on spatial and energy coordinates -- e.g., {\tt DRAGON}~\citep{2018JCAP...07..006E}, {\tt GALPROP}~\citep{2022ApJS..262...30P} and {\tt PICARD}~\citep{2014APh....55...37K} --, via stochastic differential equation based Monte Carlo simulations (e.g., {\tt CRPROPA}, \citealt{2025CoPhC.31109542M}), or from semi-analytical solutions in simplified geometries based on Green's functions, series expansion, etc. (e.g., the LBM and 1D-DM in \usine{}\footnote{\url{https://dmaurin.gitlab.io/USINE/}}, \citealt{2020CoPhC.24706942M}). 
Once $N$ is calculated for all CR isotopes, the (nearly isotropic) CR interstellar (IS) flux is given by 
$\psi^{\rm IS}=N(v/4\pi)$. To compare the modelled fluxes to the measured ones, Solar modulated top-of-atmosphere (TOA) fluxes are calculated from IS ones, using the force-field approximation \citep{1967ApJ...149L.115G,1968ApJ...154.1011G}. The energy is transformed according to
\begin{equation}
    E_{\rm k/n}^{\rm TOA}=E_{\rm k/n}^{\rm IS}-\phi_{\rm FF}\times (Z/A)\,,
    \label{eq:def_ETOA_EIS}
\end{equation}
with $Z$ and $A$ the atomic and mass number of the species, $\phi_{\rm FF}$ the solar modulation potential varying from $\sim 500$\,MV (Solar minimum) to 1200\,MV (Solar maximum). The flux at the top of the atmosphere is related to the interstellar flux by $\psi^{\rm TOA}(E^{\rm TOA}) = \psi^{\rm IS}(E^{\rm IS})\times (p^{\rm TOA}/p^{\rm IS})^2$. In the above equations, $p$ is the momentum, $E$ is the total energy, and $E_{\rm k/n}=E_{\rm k}/A$ is the kinetic energy per nucleon, with $E_{\rm k}=E-mc^2$; the rigidity, $R=pc/Ze$, is also used later on in the paper.

\subsection{Specifics of EC decay in GCRs}

In spontaneous $\beta$ decay, a neutron (resp. proton) is converted into a proton (resp. neutron) in a nucleus $X$,
\[
^A_ZX \; \stackrel{\beta^-}{\to} \;^{\;\;\;\,A}_{Z+1}X + e^- +\overline{\nu_e}\,,
\]
\[
^A_ZX \; \stackrel{\beta^+}{\to} \; ^{\;\;\;\,A}_{Z-1}X + e^+ + \nu_e\,.
\]
In contrast, EC decay is a process in which the nucleus captures one of its atomic electrons,
\[
^A_ZX + e^- \; \stackrel{EC}{\to} \;^{\;\;\;\,A}_{Z-1}X +  \nu_e\,,
\]
usually from the innermost shell ($K$-shell), which has the highest probability to interact with the nucleus.
As a result, EC decay is possible only if it is preceded by electron attachment, and only if electron stripping is not faster than the decay time. This additional decay channel thus results from a balance (or competition) between the three rates at play (\Ga{}, \Gs{} and \GEC{}), where the first two processes occur in the ISM (where electrons are available), while the last one can occur anywhere. For a CR species $i$, the first two rates are defined as
\begin{equation}
   \Gamma^{\rm att}_i=\sum_{t\,\in\,\rm ISM} \, n_t \, v_i \, \sigma^{\rm att}_{i+t}  \quad {\rm and} \quad  \Gamma^{\rm str}_i= \sum_{t\,\in\,\rm ISM}n_t \, v_i \, \sigma^{\rm str}_{i+t}\,, 
   \label{eq:gamma_att_str}
\end{equation}
where \siga{} and \sigs{} are the cross-sections for the attachment and stripping of electrons in the ISM (see~\ref{app:sig_att_str}).

As a consequence of the energy and mass dependence of the latter quantities, ions in GCRs are usually fully-ionised.
Indeed, stripping is dominant over a few tens of MeV/n and for light nuclei (see Sect~\ref{sec:timescales}), so that EC decay is suppressed. This explains why EC decay can be used to measure the time elapsed between nucleosynthesis and acceleration: species at rest in the ISM have electrons attached (EC decay enabled), but lose them once they are accelerated (EC decay blocked). Furthermore, if reacceleration occurs during the transport, the bulk of GCRs can gain a few hundreds of MeV/n, moving from a situation where EC decay is favoured (electron attachment is possible), to one where it is disfavoured (ions are mostly fully ionised).

EC decay is the sole decay mode in proton-rich nuclei, if the binding energy difference between the initial and final states is below $2m_e\simeq1022$\,keV. Otherwise, EC and $\beta^+$ are competing processes, although their respective branching ratios are not always experimentally available~\citep{2021ChPhC..45c0001K}. 
Table~\ref{tab:EC clocks} lists all nuclei undergoing pure EC decay as extracted from \textsc{Nubase}~\citep{2021ChPhC..45c0001K}, along with their half-lives, final stable daughters and, where available, their multistep decay chains.

The modification of the transport Eq.~\eqref{eq:transport_gen} to account for EC decay, along with the explicit solutions for the LBM and 1D-DM are detailed in Sect.~\ref{sec:multi-level-eq} and\textbf{}~\ref{app:solutions}.

\subsection{Dominant processes and timescales}
\label{sec:timescales}

We show in this section the timescales of the processes involved in 1D-DM, consisting of a thin disc of half-width $h$ (source and gas) and a diffusive halo of half-width $L$ (in an infinite plane). The interplay between these timescales shapes the isotopic and elemental GCR fluxes. The escape time from the diffusive volume (or diffusion time) is $\Td= L^2/(2D)$. The inelastic, attachment and stripping times in the gaseous disc are obtained from the inverse of the rates defined in Eqs.~\eqref{eq:gamma_nuc} and~\eqref{eq:gamma_att_str}; the decay time is the inverse of the decay rate Eq.~\eqref{eq:gamma_decay}.

\begin{table}[t]
\centering
\caption{Typical energy, CR charge (or atomic number) and mass number dependence} of timescales in the 1D-DM (see~\ref{app:solutions}) in the relativistic regime ($\beta\approx 1$): diffusion (diff), inelastic interactions (inel), electron radiative attachment (ra) and non-radiative attachment (nra), electron stripping (str), and $\beta$ decay; see \ref{app:sig_att_str} for a full discussion of the attachment and stripping cross-sections.
\label{tab:timescales}
\addtolength{\tabcolsep}{-2mm}
\begin{tabular}{ll}
\hline\hline
\multicolumn{2}{c}{Rates} \\ \hline
&\\[-3mm]
\Td & $\propto E^{-0.5}$  \\
\Ti & $\propto A^{-2/3}$  \\
\Ta & $\propto E\,Z^{-5}$ (ra)   \\
    & $\propto E\,Z^{-5}$ (nra)   \\
\Ts & $\propto Z^{2} / \ln E$     \\
\TEC& $\propto E$ \\
\hline
\end{tabular}
\end{table}

\begin{figure*}[t]
\includegraphics[width=\textwidth]{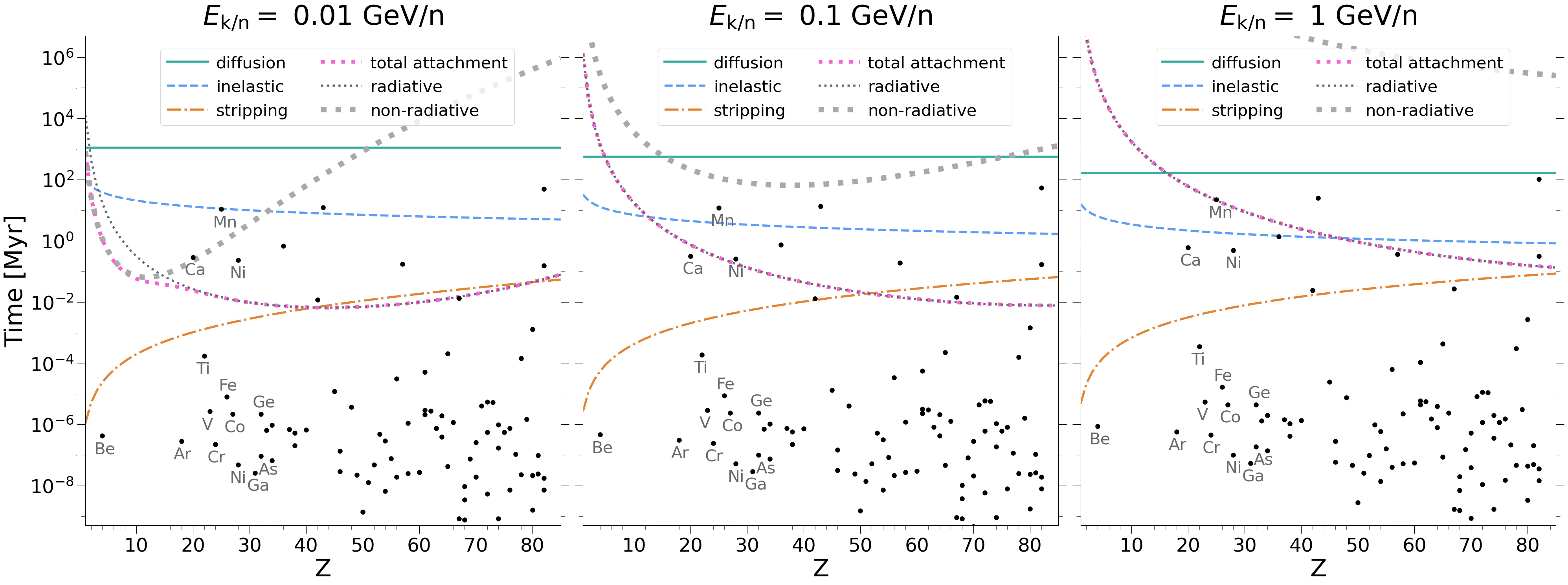}
\centering
\caption{Timescales for the processes listed in Table~\ref{tab:timescales}, as a function of the atomic number $Z$, for three energies (from left to right). The attachment time (pink dotted line) is broken down into the radiative (thin-dotted grey line) and non-radiative (thick-dotted grey line) contributions, see~\ref{app:sig_att}. The black dots represent the effective decay time, $\TEC =\gamma\tau_{\rm EC}$,} for each EC-unstable isotopes reported in Table~\ref{tab:EC clocks} (highlighted species are compared to CR data in Sect.~\ref{sec:sensibility_data}).
\label{fig:timescales}
\end{figure*}

\paragraph{Energy range, transport and destruction timescales}
We gather in Table~\ref{tab:timescales} the most salient energy, atomic number or mass number dependence of these timescales. In Fig.~\ref{fig:timescales}, we show them as a function of the atomic number $Z$, for three energies. We choose the minimum energy 10\,MeV/n (left panel) as it corresponds to the lowest IS energies measured by CR experiments (see Sect.~\ref{sec:sensibility_data}), and 1\,GeV/n as the maximum energy (right panel), because EC decay is blocked above\footnote{In passing, this is a striking difference between EC and spontaneous decay, as the former is blocked when $\Ta \gg \min(\Td,\,\Ti)$, whereas the latter is blocked when $\gamma t_{1/2}\gg \Td$.}. 
As already reported in the literature \citep[e.g.,][]{2019A&A...627A.158D}, below a few GeV/n, destruction via inelastic interactions (blue dashed line) is significant over diffusion (green solid line) for heavy nuclei (i.e.,  $\Ti<\Td$ for $Z\gtrsim10$). Actually, as discussed in \citet{2022FrP....1058841V}, diffusion prevails at increasingly larger energies for growing CR masses.

\paragraph{Attachment and stripping timescales}
The total electron attachment time (pink dotted line) is dominated by radiative attachment (thin grey dotted line): the non-radiative one (thick grey dotted line) is negligible, except for light nuclei below tens of MeV/n (left panel). Attachment is more efficient for heavy species ($\Ta\propto Z^{-5}$), while this is the opposite for stripping ($\Ts\propto Z^2$), see Table~\ref{tab:timescales}. Attachment is also more efficient at low (10\,MeV/n) than at high energy (1\,GeV/n), whereas the energy dependence for stripping is milder. Actually, in the sub-GeV range, the energy dependence is non-trivial for these two processes. Overall, we have $\Ta<\Ts$ for heavy nuclei and low energy only, which is the optimal regime to ensure at least one electron can be attached. 

\paragraph{Decay time}
EC decay is possible if an electron is attached, that is, if $\Ta<\min(\Td,\,\Ti)$, which is the case at low energy and for heavy nuclei. The majority of the known EC isotopes have $\TEC\ll\Ts$ (black dots below the orange dash-dotted line). For these short-lived CRs, decay occurs immediately after attachment. For the remaining CRs (i.e., with $\TEC \gtrsim 10^{-4}$\,Myr), the interplay among stripping, attachment and decay times has to be taken fully into account to determine their effective decay rate. We refer readers to \citet{1984ApJ...279..144L, 1984ApJS...56..369L, 1985Ap&SS.114..365L} for a complementary analysis of these behaviours (carried out in the LBM).

\section{Impact of EC-decay, uncertainties, and validity}
\label{sec:Results}


\subsection{Multi-level transport equation}
\label{sec:multi-level-eq}

The transport equations, for a $\beta$-unstable species and a generic source term $Q$, in the LBM and thin-disc 1D-DM, are given respectively by
\begin{equation}
    \frac{N}{\tau_{\rm esc}} + \left(\Gi + \Gb\right)N = Q\,,
\label{eq:LBM}
\end{equation}
and
\begin{equation}
    -D\frac{d^2N}{dz^2} + \left(2h \delta(z) \,\Gi + \Gb\right)N = 2h \delta(z) \,Q\,.
    \label{eq:DM}
\end{equation}

The analytical solutions of the transport equation, for stable and $\beta$-radioactive GCR ions, are gathered in \cite{2020CoPhC.24706942M}, for the LBM and 1D-DM.
We note that the source term cancels out in all IS flux ratios computed in Section~\ref{sec:Results} and Section~\ref{sec:sensibility_data}; consequently, it is not included in the model.
In this work, we extend these solutions to the EC case, without energy redistribution and without galactic wind (for the 1D-DM), which allows us to use the simple analytical solutions reported in~\ref{app:solutions}.

Energy losses shape GCR spectra below hundreds of MeV/n, roughly shifting downwards their energy: GCRs at $E_{\rm k/n}$ actually come from $E'_{\rm k/n}>E_{\rm k/n}$, that is, an energy where EC decay is less efficient and impacting (see previous section). Consequently, the {\em no-energy loss} calculation overestimates the impact of EC decay. 
We explore below several aspects of EC decay: validity of some approximations, impact of uncertainties, detectability of EC decay in current CR data (Sect.~\ref{sec:sensibility_data}). We will show that our conclusions, obtained in the {\em no-energy loss} framework, are conservative, that is, they would be even more valid if energy losses were to be accounted for.

Compared to the standard solutions, we now have to follow the various ionisation states of the GCR species to include EC decay. Indeed, whereas CRs are expected to be fully ionised at high-enough energy (see Fig.~\ref{fig:timescales} in Sect.~\ref{sec:timescales}), several ionisation states may co-exist at lower energy \citep{1984ApJS...56..369L,1985Ap&SS.114..365L}.
As a result, instead of solving the transport equation for the differential density of the CR ion $N$, we have to solve it for a tower of ionised states, denoted below $N_0$ (fully ionised), $N_1$ (one $e^-$ attached), $N_2$ (two $e^-$ attached), etc. Each level is connected to the one below and above via the electron attachment and stripping rates, \Ga and \Gs, the decay rate $\GEC = 1/(\gamma \tau_{EC})$ being enabled for all $N_{j>0}$ (i.e., at least one electron attached). The total measured flux of this CR species is then given by 
\begin{equation}
N_{(n+1)~{\rm lev}}=N_0+N_1+N_2+\dots+N_n\,,
\label{eq:def_Nnlev}
\end{equation}
where $n\leq Z$ (with $Z$ the charge of the CR under scrutiny) is the enforced truncation level in the calculation.

Throughout this section, our reference calculation (with EC decay) is a 2-level model, denoted \Ntwolev{}, as considered in the literature (i.e., either fully ionised CR or with a single electron attached). We use the 1D-DM and the best-fit transport parameters of \citet{2024A&A...688A..17F}, where neither convection nor reacceleration is favoured by CR data.

\subsection{Impact of EC decay on isotopic IS fluxes}

\begin{figure*}[!th]
\centering
\includegraphics[trim={0 3.5cm 0 0},clip,width=\textwidth]{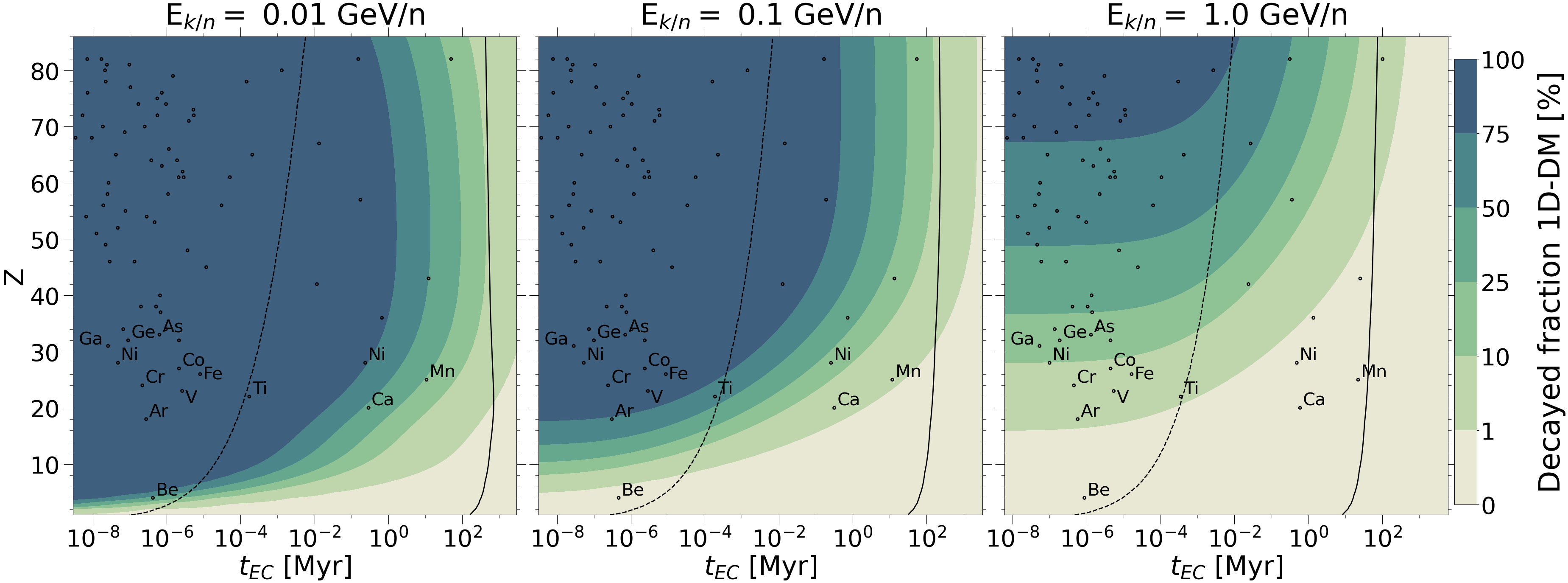}
\includegraphics[trim={0 0 0 2.7cm},clip,width=\textwidth]{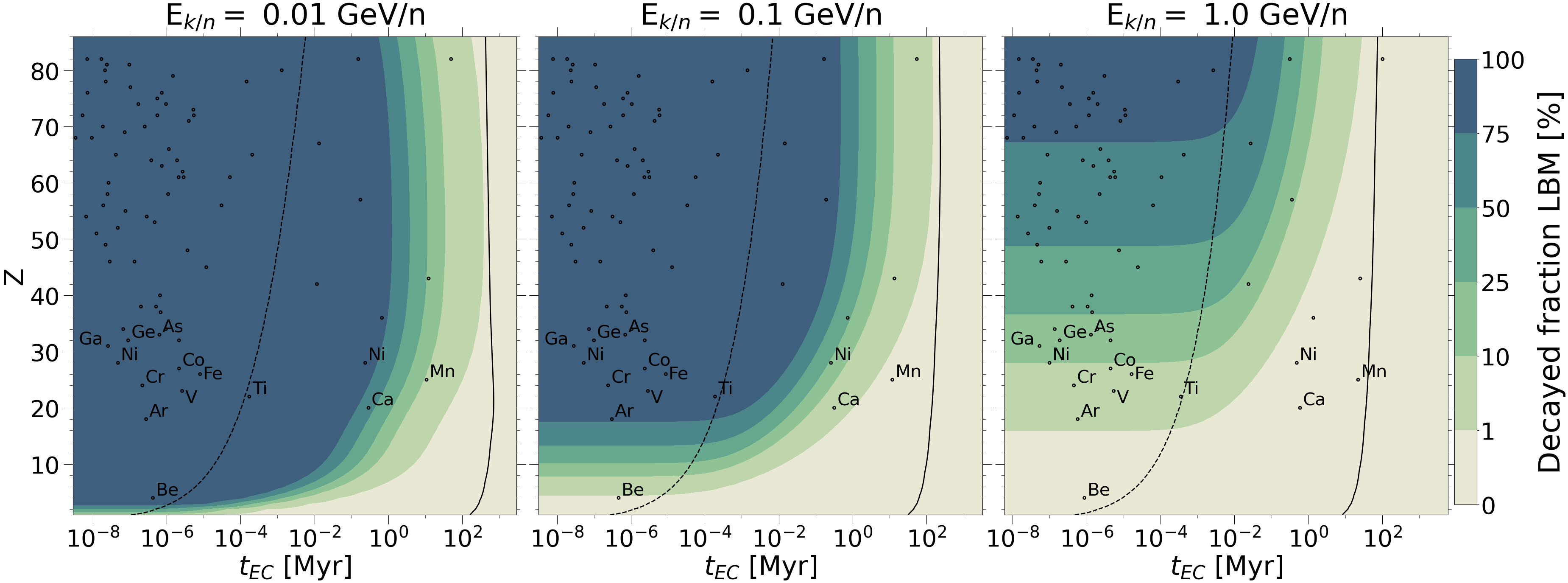}
\caption{Colour-coded EC-decayed fraction, Eq.~\eqref{eq:dec_frac}, in the $\TEC$--$Z$ plane for the 1D-DM (top) and LBM (bottom), for three IS energies. The two near-vertical lines separate short-lived  ($\TEC < 0.1 \,\Ts$), intermediate-lived ($0.1 \,\Ts < \TEC < 10 \, \Td $), and long-lived isotopes ($\TEC > 10 \, \Td$), where $\TEC=\gamma\tau_{\rm EC}$}. Symbols indicate {known} EC-unstable isotopes (see Table~\ref{tab:EC clocks}), highlighting the names of those for which CR data exist (see Sect.~\ref{sec:sensibility_data}).
\label{fig:dec_frac_IS}
\end{figure*}

\paragraph{Impact in the 1D-DM}
The top panels of Fig.~\ref{fig:dec_frac_IS} show the decayed fraction,
\begin{equation}
    \label{eq:dec_frac}
    f = 1 - \frac{\Ntwolev}{\Nonelev}= 1 - \frac{\Ftwolev}{\Fonelev}\,,
\end{equation}
as a function of a generic $\TEC$ and $Z$. In the above equation, \Ntwolev, defined in \eqref{eq:def_Nnlev}, corresponds to the calculation tracking only two states (fully ionised, and single electron attached), as used in the literature, whereas \Nonelev{} is the calculation without EC decay (only the fully ionised state is enabled).

As expected from the timescales, EC decay has a maximum impact on isotopic fluxes at low energy (left panel), and mostly at high $Z$ and short decay times (top-left corners in the various panels). 
We stress that above a few GeV/n (not shown), the decayed fraction becomes less than 1\% for EC-unstable species currently measured in CR data (those with their element name in the plot); above tens of GeV/n, this applies to all EC-unstable isotopes.

The near-vertical line on the right-hand side of all panels corresponds to $\TEC (= \gamma\tau_{\rm EC}) = 10 \, \Td$: isotopes on the right-hand side of this line mostly escape from the Galaxy before having time to decay, so that their decayed fraction $f<1\%$. The other near-vertical line (on the left) corresponds to the condition $\TEC = 0.1 \,\Ts$: isotopes on the left-hand side of this line (i.e., where most EC-decay isotopes lie) decay immediately after they attach an electron, hence, their  decay is solely controlled by the attachment time. As a result, the decayed fraction is $f\simeq 100\%$ for all species at low energy (left panel), whereas at higher energy (central and right panel), it goes from a few percent up to 100\% for growing $Z$, following the behaviour of the attachment time.
In the in-between region (i.e., intermediate-lived isotopes), the decay impact depends on the interplay between the EC decay, stripping, and attachment times, and can go from 100\% to less than 1\%, depending on the energy and the charge of the isotope.

\begin{figure*}[!th]
\centering
\includegraphics[width=\textwidth]{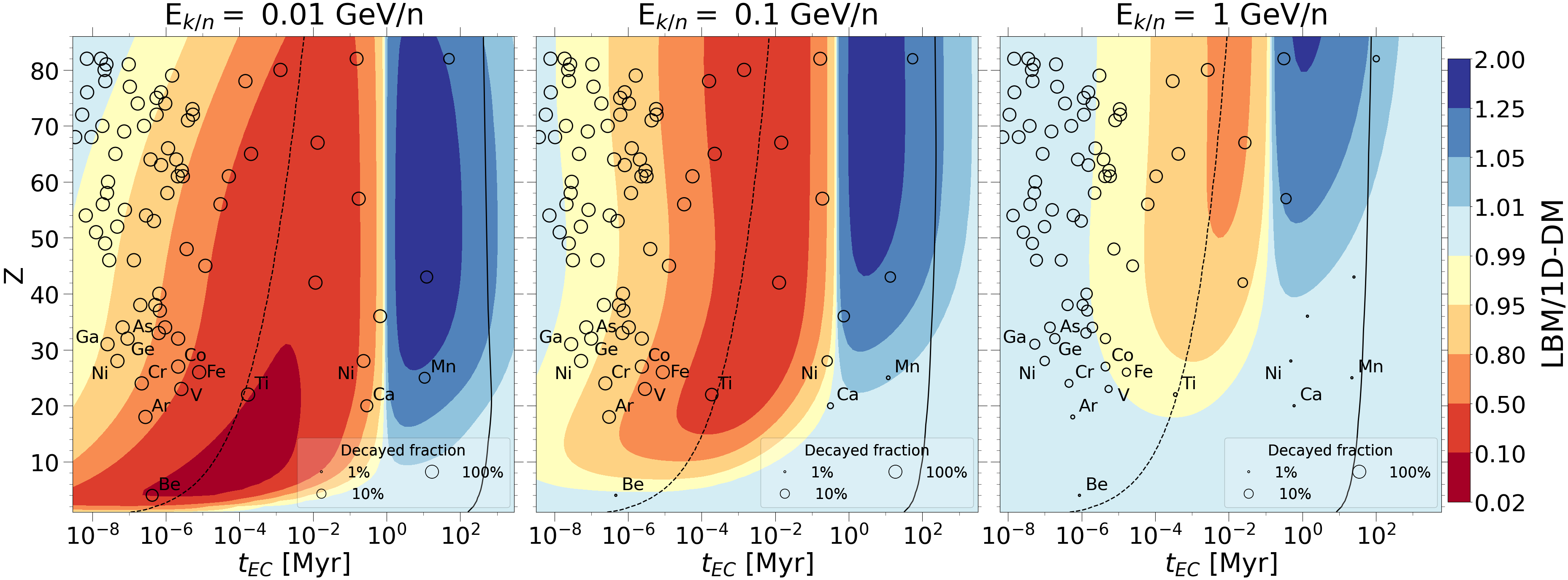}
\caption{Same axes and configuration as in Fig.~\ref{fig:dec_frac_IS}, but now the colour scale shows the ratio of the LBM to the 1D-DM calculation, defined in Eq.~\eqref{eq:def_ratio}. The size of the circles encodes the decayed fraction $f$ of the EC-unstable isotopes.}
\label{fig:dec_frac_IS_LMB_1DDM}
\end{figure*}

\paragraph{Differences between the LBM and 1D-DM}
Studies on $Z>30$ species have mostly been carried out in the context of LBM, so we extend our discussion to this model. We first stress that, for stable species, the solution of the LBM and the 1D-DM are formally similar, provided that \citep[e.g.,][]{2001ApJ...547..264J}
\begin{equation}
    \tau_{\rm esc} = \frac{hL}{D}\,.
    \label{eq:equiv_LBM_DM}
\end{equation}
We use the above relation in our calculations, in order to have meaningful comparisons between the two models. However, the 1D-DM is not reducible to the spatially homogeneous LBM for $\beta$-unstable species \citep{1975Ap&SS..32..265P}, and we expect the same irreducibility for EC-unstable species.
Indeed, this non-equivalence is directly seen from the different formal solutions, Eqs.~\eqref{eq:Sol2lev_LBM} for the LBM and~\eqref{eq:Sol2lev_DM} for the 1D-DM, where the LBM solution is more dominated by the decay term than is the 1D-DM one.

The decayed fraction in the LBM is shown in the bottom panel of Fig.~\ref{fig:dec_frac_IS}. Qualitatively, the trend is very similar in the 1D-DM and LBM (compare the top and middle panels). However, quantitatively, the decayed fraction shows significant differences for intermediate-lived isotopes (those between the two vertical lines). A complementary view is provided by forming the ratio of the LBM to the 1D-DM solutions (Fig.~\ref{fig:dec_frac_IS_LMB_1DDM}), 
\begin{equation}
\label{eq:def_ratio}
{\cal R}^{\rm LBM/1D-DM} = \frac{N_{2\,{\rm lev}}^{\rm LBM}}{N_{2\,{\rm lev}}^{\rm 1D-DM}} = \frac{\psi_{2\,{\rm lev}}^{\rm LBM}}{\psi_{2\,{\rm lev}}^{\rm 1D-DM}}\,.
\end{equation}
For long-lived isotopes (close or beyond the vertical line on the right), the ratio is $1$, as expected for {\em stable} species. For intermediate-lived species with $Z>40$ and $E_{\rm k/n}<100$\,MeV/n, the LBM is a factor 1.5 larger than the 1D-DM calculation, whereas just at the transition between short-lived and intermediate-lived species (near-vertical line on the left-hand side of the plot), the LBM can be half the value of the 1D-DM, or even less for $Z<30$ at very low energy (bottom left corner of the left of Fig~\ref{fig:dec_frac_IS_LMB_1DDM}).
Among the currently measured CR fluxes, $^{44}{\rm Ti}$ is the EC-stable species most impacted by the model choice.

\begin{figure*}[t]
\centering
\includegraphics[trim={0 3.5cm 0 0},clip,width=0.95\textwidth]{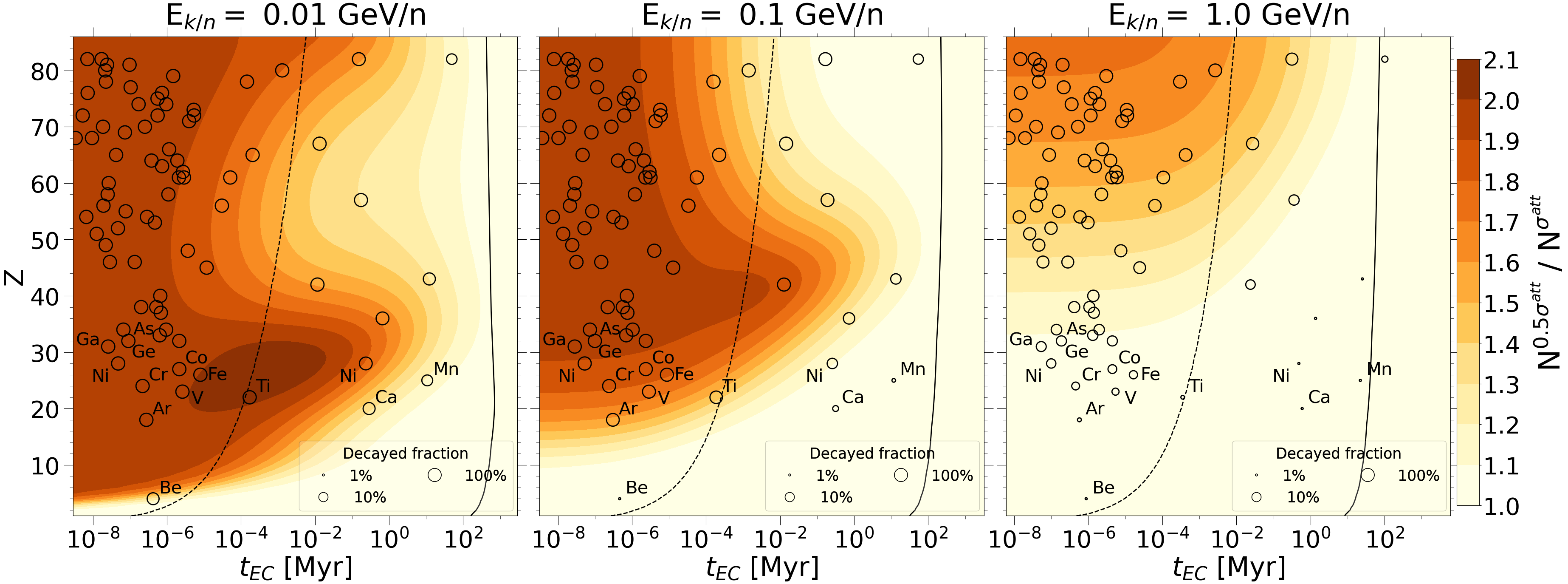}
\includegraphics[trim={0 0 0 2.7cm},clip,width=0.95\textwidth]{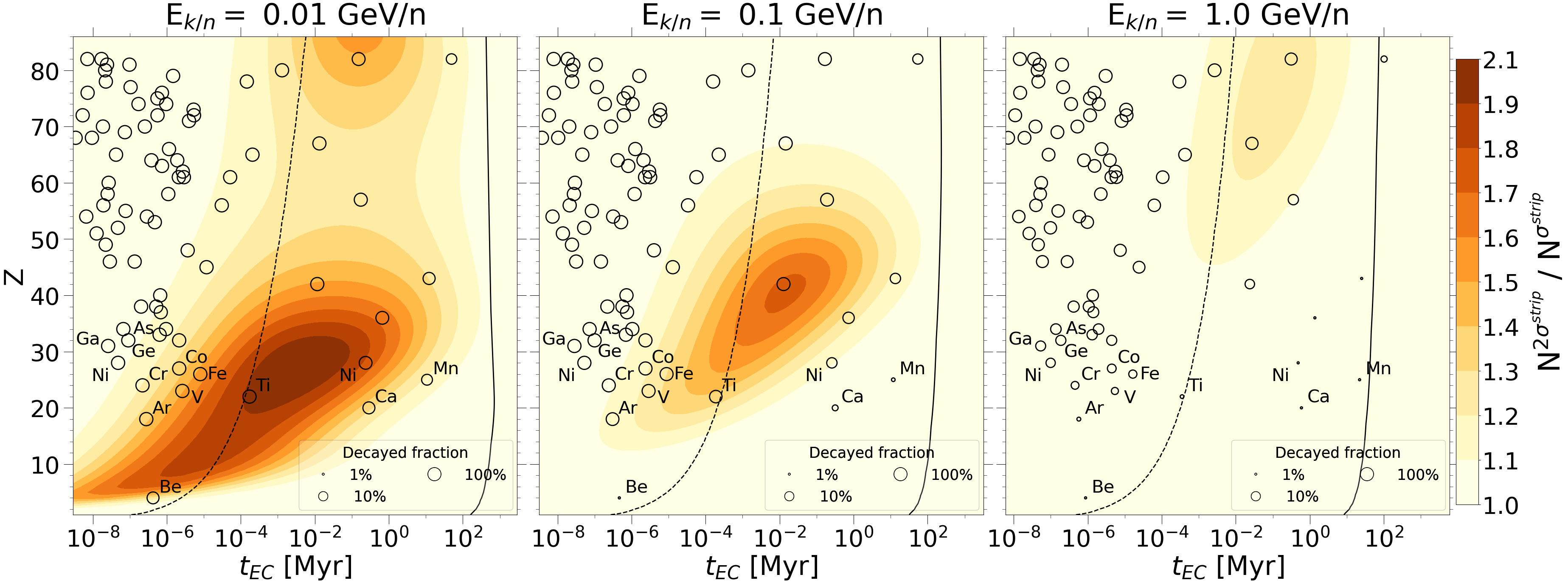}
\caption{Same as bottom panel of Fig.~\ref{fig:dec_frac_IS}, but for the ratio ${\cal R}^{\sigma_{\rm new}}$, see Eq.~\eqref{eq:def_ratio_err}, when halving the attachment (top) and doubling the stripping cross-sections (bottom).}
\label{fig:half_att_strip}
\end{figure*}

\subsection{Impact of \siga{} and \sigs{} uncertainties}
\label{sec:impact_att_str_uncertainty}

As discussed in~\ref{app:sig_att_str}, the attachment and stripping cross-sections (\siga{} and \sigs{}) used in the GCR literature are based on data and parametrisations dating back from the 1970s~\citep{osti_6508366, 1979PhDT........67C, 1984ApJS...56..369L}. Moreover, these formulas are derived in the context of single electron attachment, and their extrapolation to higher ionised states could introduce additional uncertainties in the models (see~\ref{sec:sigma_ext}). Therefore, it is important to quantify the impact of a variation in $\siga$ and $\sigs$ on the isotopic fluxes, and for that, we focus on the ratio (in the 2-level configuration)
\begin{equation}
    \label{eq:def_ratio_err}
    {\cal R}^{\sigma_{\rm new}}=\frac{N_{2\,{\rm lev}}^{\sigma_{\rm new}}}{N_{2\,{\rm lev}}^{\sigma_{\rm ref}}} = \frac{\psi_{2\,{\rm lev}}^{\sigma_{\rm new}}}{\psi_{2\,{\rm lev}}^{\sigma_{\rm ref}}}\,.
\end{equation}

In Fig.~\ref{fig:half_att_strip}, this ratio is shown for a halved \siga{} (top) and a doubled \sigs{} (bottom). This choice allows us to show the two cases with the same colour scale, easing a visual comparison and discussion (of the impact of a factor 2 uncertainty on these cross-sections). Indeed, both changes lead to less decay, hence a larger $N^{\sigma_{\rm new}}$ and ${\cal R}^{\sigma_{\rm new}}>1$, as it takes a longer (resp. shorter) time to attach (resp. strip) an electron, slowing down and decreasing the decay probability.  We stress that if, instead, we were to double \siga{} and half \sigs{}, this would lead to the opposite behaviour -- more decay, smaller $N^{\sigma_{\rm new}}$, and ${\cal R}^{\sigma_{\rm new}}<1$ --, while following a similar pattern in terms of the dependence with $E_{\rm k/n}$, $\TEC$ and $Z$.

\paragraph{Uncertainties on $\siga$}
In the top panel, we observe a growing impact of EC decay towards lower energies (i.e., from right to left panels), and going from intermediate- to short-lived isotopes for larger $Z$ (upper-left corners of the plots). Long-lived isotopes are insensitive to the modified cross-section, i.e., ${\cal R}^{0.5\siga}\!(\TEC\gtrsim1\,{\rm Myr})\!=\!1$. The impact of halving \siga{} (top panel) is important for short-lived nuclei (${\cal R}^{0.5\siga}\!\!=2$), as attachment directly drives the decay time. The region of intermediate-lived nuclei -- near the vertical line on the left-hand side -- is also impacted, as it is very sensitive to the balance between attachment, stripping, and decay. 
Looking back at the timescales in Fig.~\ref{fig:timescales}, among the species with CR data (i.e., highlighted names in the figure), the curve for \Ta{} with a halved decay time would get further away from $^{53}$Mn and $^{41}$Ca, but closer to $^{44}$Ti, $^{55}$Fe, etc. This implies a maximum impact for the latter, where the decayed flux is changed by more than a factor two at low energy (top left panel in Fig.~\ref{fig:half_att_strip}).

\paragraph{Uncertainties on $\sigs$}
A comparison of the bottom and top panels of Fig.~\ref{fig:half_att_strip} shows that changing the stripping and attachment times have very different impacts on the decaying fractions. In addition to have no effect on long-lived isotopes and high energies, there is also no effect on short-lived nuclei, as their decay is not limited by stripping (left side of the plots). Intermediate-lived nuclei are impacted at the same level and for the same nuclei as when halving \siga{} (compare the top and bottom panels), owing to a similar modification of the balance between attachment, stripping, and decay. This leads to ${\cal R}^{2\sigs}\!\!=2$ at 10\,MeV/n (left panel). It decreases at higher energies (right panels), while the most impacted nuclei are shifted towards larger $Z$ and $\TEC$.

\paragraph{Overall impact of these uncertainties}
A last step must be taken to conclude whether more precise cross-sections are needed or not for \siga{} ad \sigs{}. Indeed, the impact of cross-section uncertainties (${\cal R}^{\sigma_{\rm new}}$) must be considered in the light of the overall EC-decayed fraction. For existing isotopes, the latter is indicated in Fig.~\ref{fig:half_att_strip}  by the size of the circles (or, equivalently, encoded in the $f$ values shown in the top panels of Fig.~\ref{fig:dec_frac_IS}):
\begin{itemize}
    \item in the bottom-right region of the $\TEC$--$Z$ plane, we have ${\cal R}^{\siga_{\rm new}}\simeq{\cal R}^{\sigs_{\rm new}}\simeq1$ and $f\lesssim 1\%$. As there is no foreseeable scenario in which CR data will reach the percent precision on isotopes, large uncertainties on \siga{} and \sigs{} (up to a factor two or more) are non-impacting.
    \item in the top-left region of the $\TEC$--$Z$ plane, we have ${\cal R}^{\siga_{\rm new}}\gtrsim2$, ${\cal R}^{\sigs_{\rm new}}\approx1$ and $f\sim 100\%$. The latter number means that EC-decay decreases the fluxes by a factor $\gtrsim 100$ and is insensitive to uncertainties on \sigs{}, but that a factor 2 uncertainty on \siga{} translates to a factor of two uncertainty on this suppression factor: while significant, this does not qualitatively change the conclusions regarding the detectability of EC-decay in CR data (the data precision goes from a few percent to tens of percent, see Sect.~\ref{sec:sensibility_data}).
    \item intermediate-lived nuclei (inner region between the two near-vertical lines) with $Z\approx30$ (i.e., $^{44}$Ti, $^{55}$Fe, etc.): this is a region, at low-energy, where both ${\cal R}^{\siga_{\rm new}}$ and ${\cal R}^{\sigs_{\rm new}}$ are $\gtrsim2$, and $5\lesssim f\lesssim 50\%$ (depending on the energy). Consequently, the cross-section uncertainties maximally impact the conclusions for this group of nuclei: the EC-decay impact can go from very small (few percent) to very significant (up to a factor 2), compared to the CR data precision.
\end{itemize}

We stress that we obtain similar results for the LBM, and also for multi-level models (not shown). We will come back to these uncertainties in Sect.~\ref{sec:sensibility_data}. However, we can already say that, overall, most EC-unstable isotopes are insensitive to large uncertainties on \sigs{}, while they show no qualitative change due to uncertainties on \siga{}. Only the flux of isotopes with $\TEC\approx{\cal O}(\rm kyr)$ at IS energies below a few hundreds of MeV/n is critically sensitive to uncertainties on both cross-sections.

\begin{figure*}[t]
\centering
\includegraphics[trim={0 3.5cm 0 0},clip,width=0.95\textwidth]{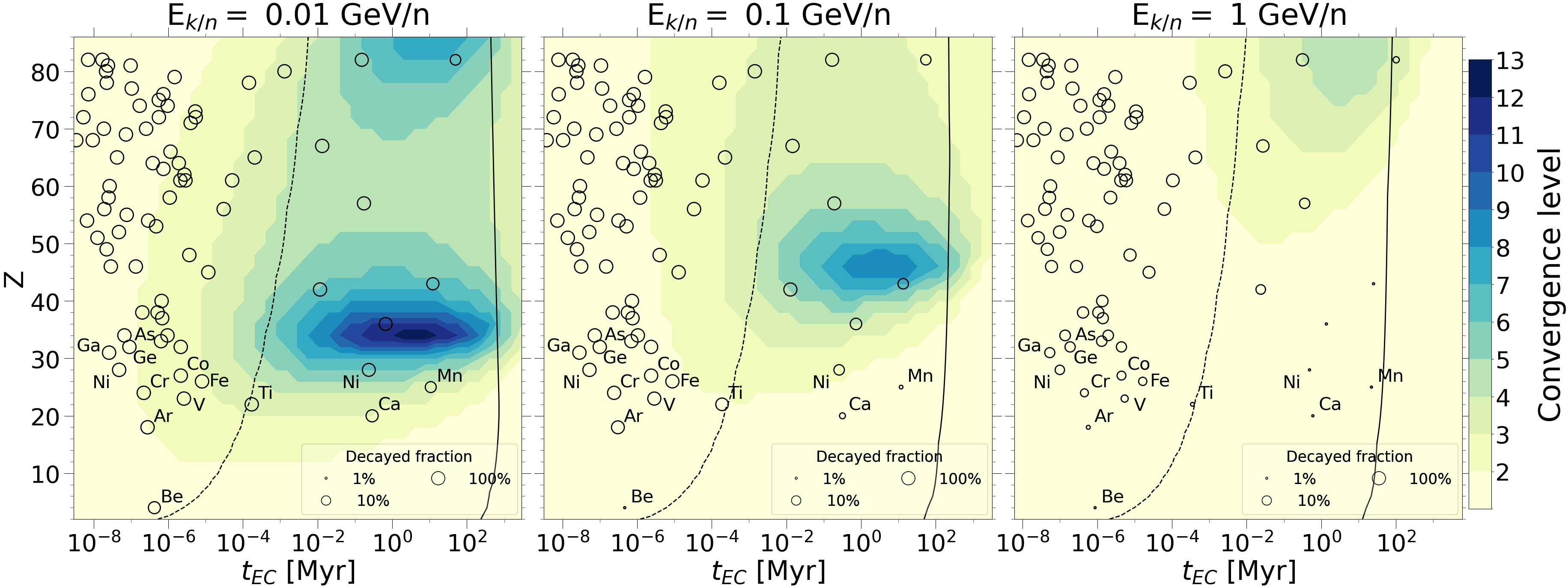}
\includegraphics[trim={0 0 0 2.7cm},clip,width=0.95\textwidth]{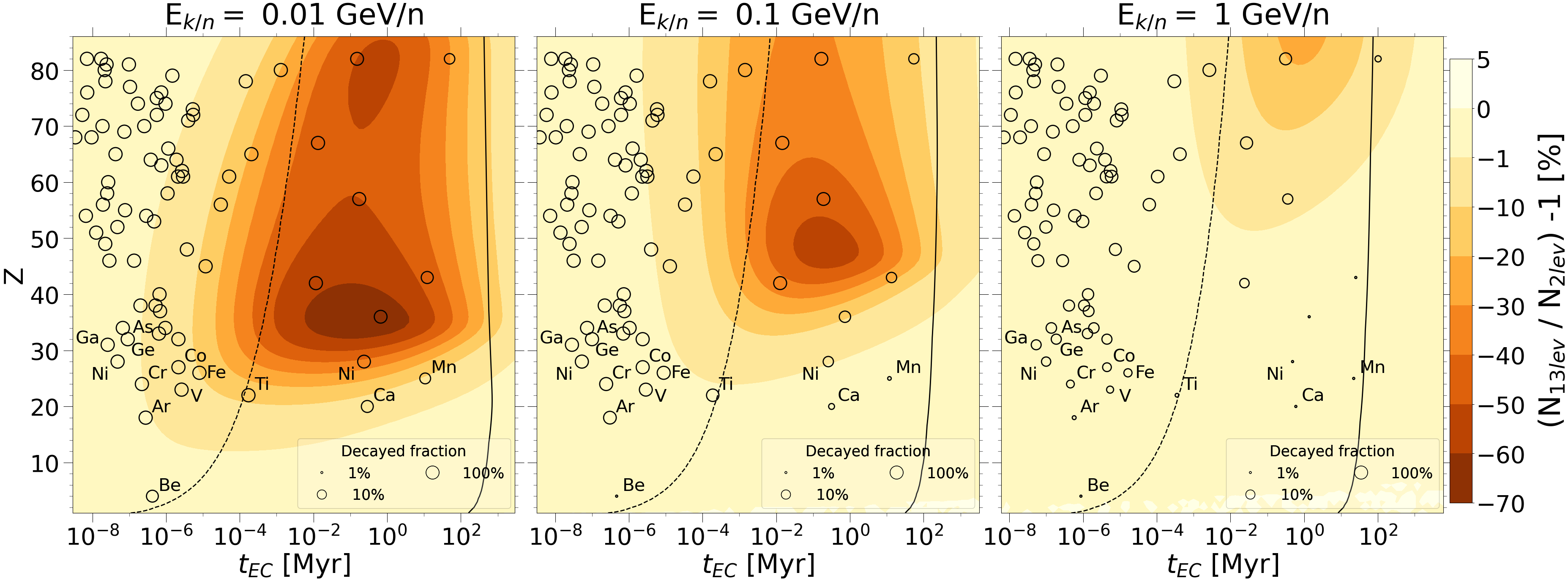}
\caption{Same as in previous figures, but showing $n$ at which convergence is reached (top row), and the relative difference between the literature approximation $n=2$ and the converged solution (bottom row); both calculations are for the 1D-DM.}
\label{fig:solcomp_multi}
\end{figure*}

\subsection{Validity of currently used approximations}
\label{sec:res_validityapprox}

Previous studies \citep{1985Ap&SS.114..365L, 1998A&A...336L..61S, 2001SSRv...99...41C, 2001AdSpR..27..737J, 2003ApJS..144..153W, 2005PhDT.........6S} rely on two assumptions. First, partially ionised nuclei have different charges, so that all CR species (stable ones included) should be differently affected by diffusion. These works do not explicitly state whether they account for this.
Second, GCR calculations do not consider more than one electron attached, because the corresponding populations drop quickly~\citep{1979PhDT........67C, 1984ApJS...56..369L}. Experimentally, various ionic states, up to hundreds of MeV/n, have been observed in the Solar neighbourhood. A GCR origin for high-ionic Fe states was even briefly considered~\citep{1990ApJ...359L...5B, 1995ApJ...438L..83T}. However, the presence of ionic states is only firmly established for transient Solar energetic particles~\citep{1995ApJ...444L.109T, 2007SSRv..130..273K, 2018SSRv..214...61R} and anomalous CRs~\citep{1993ApJ...411..418D, 1996ApJ...466L..43M, 2007BRASP..71..980B}.

\paragraph{Accounting for an effective charge in the transport}

Different ionised states of the same isotope have different charges. As diffusion depends on the rigidity, one should in principle account for this charge difference. We checked that the solution with and without this approximation differs by $\lesssim 1\%$ at all energies and for all species (not shown). This validates the use of the same charge for the transport of all ionic states of the same isotope.

\paragraph{Multi- vs 2-level calculation}

As defined in Eq.~\eqref{eq:def_Nnlev} and \ref{app:sol_multi-level}, the calculation with $n$ levels considers CR ions with zero, one, two, etc., up to $n-1$ electrons attached. We define the quantity $\Delta_{(n)}$,
\[
\Delta_{(n)} = \left| \frac{N_{n\,{\rm lev}}}{N_{(n-1)\,{\rm lev}}} - 1 \right|\,,
\]
to track the relative difference between the calculation carried out up with $n$ and $n-1$ levels.
We consider to have reached the asymptotic solution (or convergence) at $n$ if $\Delta_{(n)} \leq 1\%$. 

The top panel of Fig.~\ref{fig:solcomp_multi} shows $n$ at which the convergence is reached: above a few hundreds of MeV/n (IS), the 2-level approximation is sufficient for all EC-unstable species, but for a few nuclei with $Z\gtrsim40$ and $\TEC\gtrsim10$\,kyr. Nevertheless, at 10 MeV/n, the convergence is reached for $n=13$ levels for intermediate-lived nuclei of charge $Z=30-40$. For the LBM, similar conclusions are reached (same number of levels to reach convergence), though with slightly tighter contours along the$\TEC$ axis.

The bottom panel shows the relative difference between the converged solution (assumed to be $N_{13\,{\rm lev}}$) and $N_{2\,\rm lev}$. The flux is further suppressed when extra levels are present, and the largest for regions of large $n$ (compare the top and bottom panels). The relative difference between $N_{3\,\rm lev}$ and $N_{2\,\rm lev}$ (not shown) indicates that a large fraction of the extra-suppression is already contained in the $n=3$ level calculation, whereas each new level has a decreasing impact. This is not surprising, because going from 1 to 2 attached electrons doubles the EC-decay rate (two instead of one electron are available in the $K$-shell), whereas supplementary electrons (in other shells) do not enhance the decay and come at the cost of multiplying the overall attachment time.

Among the isotopes for which CR data exist (i.e., those with the names highlighted in the plot), the converged solution differs at most by $\lesssim 20\%$ (apart from $^{59}$Ni showing a difference $\sim 37\%$), and only for very low energy (bottom left panel). More importantly, in the range targeted by TOA current experiments, that is, above several hundreds of MeV/n in equivalent IS energies (in-between the middle and right panels), the difference in the decayed fraction does not even reach the percent level, validating the use of the 2-level approximation. Similar conclusions are reached for the LBM.

\begin{table}[t]
\small
\centering
\caption{CR isotopic fraction extracted from the following TOA data in \crdb{}: $^{7}{\rm Be}$/Be \citep{1977ApJ...212..262H, 1977ApJ...217..859G, 1977ApL....18..125W, 1978ApJ...226..355B, 1979ICRC....1..389W, 1980ApJ...239L.139W, 1981ICRC....2...72G, 1998ApJ...501L..59C, 2001ICRC....5.1751C, 2021Univ....7..183N}, $^{37}{\rm Ar}$/Ar \citep{1981ApJ...246.1014Y, 1985ICRC....2...88W, 2009ApJ...695..666O}, $^{41}{\rm Ca}$/Ca \citep{1981ApJ...246.1014Y, 1985ICRC....2...88W, 1993ICRC....1..571L, 2001ICRC....5.1751C, 2009ApJ...695..666O}; $^{44}{\rm Ti}$/Ti, $^{49}{\rm V}$/V and $^{51}{\rm Cr}$/Cr \citep{1981ApJ...246.1014Y, 1981ICRC....2...80W, 1993ApJ...405..567L, 2001ICRC....5.1751C}; $^{53}{\rm Mn}$/Mn \citep{1981ApJ...246.1014Y, 1981ICRC....2...80W, 1993ApJ...405..567L, 1997ApJ...481..241D, 2001ICRC....5.1751C}, $^{55}{\rm Fe}$/Fe \citep{1981ApJ...246.1014Y, 2001AdSpR..27..773W, 2001ICRC....5.1751C}, $^{56}{\rm Ni}$/Ni \citep{1981ApJ...246.1014Y}, $^{57}{\rm Co}$/Co \citep{1993ApJ...405..567L, 2001ICRC....5.1751C}, $^{59}{\rm Ni}$/Ni \citep{2001ICRC....5.1751C};  $^{67}{\rm Ga}$/Ga, $^{71}{\rm Ge}$/Ge
and $^{73}{\rm As}$/As \citep{2022ApJ...936...13B}. 
The last column shows the variance-weighted mean, $\langle x \rangle=(\sum_i w_i x_i)/(\sum_i w_i$) with $w_i=1/\sigma_i^2$ -- which corresponds to the maximum likelihood estimator for independent and normally distributed values -- and its standard error ($\sigma_{\langle x \rangle}=\sqrt{1/(\sum_i w_i)}$, calculated over all energies of the CR data available.
We stress that the} dispersion between the data points is much larger than the weighted errors for $20\leq Z\leq26$ isotopes, and that $Z\geq 28$ isotopes have only one data point.\label{tab:CRDATA_TOA_ISOTOPIC_RATIOS}
\begin{tabular}{rcl}
\hline\hline
 $Z$ & Isotopic ratio & Weighed mean (error) \\
\hline
4      & $^{7}{\rm Be}$/Be   &  0.57 ($\pm$ 0.06) \\
18     & $^{37}{\rm Ar}$/Ar  &  0.03 ($\pm$ 0.02) \\
20     & $^{41}{\rm Ca}$/Ca  &  0.07 ($\pm$ 0.02) \\
22     & $^{44}{\rm Ti}$/Ti  &  0.01 ($\pm$ 0.003) \\
23     & $^{49}{\rm V}$/V    &  0.40 ($\pm$ 0.05) \\
24     & $^{51}{\rm Cr}$/Cr  &  0.24 ($\pm$ 0.02) \\
25     & $^{53}{\rm Mn}$/Mn  &  0.44 ($\pm$ 0.03) \\
26     & $^{55}{\rm Fe}$/Fe  &  0.04 ($\pm$ 0.02) \\
27     & $^{57}{\rm Co}$/Co  &  0.32 ($\pm$ 0.20) \\
28     & $^{56}{\rm Ni}$/Ni  &  0.04 ($\pm$ 0.03) \\
28     & $^{59}{\rm Ni}$/Ni  &  0.02 ($\pm$ 0.02) \\
31     & $^{67}{\rm Ga}$/Ga  &  0.07 ($\pm$ 0.04) \\
32     & $^{71}{\rm Ge}$/Ge  &  0.11 ($\pm$ 0.07) \\
33     & $^{73}{\rm As}$/As  &  0.36 ($\pm$ 0.23) \\
\hline
\end{tabular}
\end{table}

\section{Detectability of EC decay in current data}
\label{sec:sensibility_data}

The last step of our analysis is to assess whether the impact of EC-decay is large enough to be seen in current CR data. This impact was calculated at IS energies and for the isotopic fluxes in the previous sections. However, the most precise CR data are elemental fluxes or ratios, in which the decayed fraction is diluted by the isotopic fraction (in the element) of the EC-unstable species. In the context of this simplified study, this limits the analysis to  elemental fluxes for which the CR isotopic abundances of the EC-unstable nuclei have been measured, gathered in Table~\ref{tab:CRDATA_TOA_ISOTOPIC_RATIOS}.
Moreover, we also need to account for Solar modulation, as almost all data are taken in the Solar neighbourhood. We discuss below these various cases, based on the CR data extracted from the \crdb{}\footnote{\url{https://lpsc.in2p3.fr/crdb}} \citep{2014A&A...569A..32M, 2020Univ....6..102M, 2023EPJC...83..971M} python library\footnote{\url{https://pypi.org/project/crdb/}}.

\begin{table}[t]
\small
\centering
\caption{Detectability of EC-decay (in the 1D-DM)  with current IS CR data. The first three columns show the charge, name, and half-life of the EC-unstable species (for which IS elemental flux data exist). The next three columns show the element name, energy range, and best precision reached by current elemental IS fluxes from Voyager\,1 \citep{2016ApJ...831...18C}. We then show the expected decayed fraction computed from the 1D-DM model at 20\,MeV/n IS energy for the isotopic (next-to-last column) and the elemental fluxes (last column). The latter is calculated from the isotopic one, accounting for the isotopic fractions reported in Table~\ref{tab:CRDATA_TOA_ISOTOPIC_RATIOS}. We stress that both $^{56}$Ni and $^{59}$Ni contribute to the Ni flux.}
\label{tab:CRDATA_IS_fluxes}
\begin{tabular}{rlrlcrrr}
\hline\hline
\multicolumn{3}{c}{\!EC-unstable species} & \multicolumn{3}{c}{Elemental IS data} & \multicolumn{2}{c}{\!\!\!\!$f^{\rm IS}_{\rm dec}$ (20\,MeV/n)\!\!\!\!}\\
$Z$ &\!\!\!\!Name\!\!\!\!& $t_{1/2}$  &Qty\!\!\! &$\!\!\!\!\!\!E_{\rm k/n}$\,[MeV/n]\!\!\!\!\!\!&\!Prec.\!&\!\!\!\!(in nuc)\!\!\!\!&\!\!\!(in $Z$)\!\!\!\\
\hline
4      & $^{7}{\rm Be}$    & 53.22\,d    & Be & [61-96]   & 17\% & 3\%   & 2\% \\
18     & $^{37}{\rm Ar}$   & 35.011\,d   & Ar &\!\![117-176]\!\!& 22\% & 99\%  & 3\% \\
20     & $^{41}{\rm Ca}$   & 99.4\,kyr   & Ca & [18-193]  & 15\% & 27\%  & 2\% \\
22     & $^{44}{\rm Ti}$   & 59.1\,yr    & Ti & [75-199]  & 17\% & 95\%  & 1\% \\
23     & $^{49}{\rm V}$    & 330\,d      & V  &\!\![129-203]\!\!& 22\% & 99\%  & 40\% \\
24     & $^{51}{\rm Cr}$   &\!\!\!\!27.7015\,d  & Cr & [19-205]  & 17\% & 99\%  & 24\% \\
25     & $^{53}{\rm Mn}$\!\! & 3.7\,Myr    & Mn &\!\![153-209]\!\!& 20\% & 14\%  & 6\% \\
26     & $^{55}{\rm Fe}$   &\!\!\!\!2.7562\,yr  & Fe & [16-280]  & 10\% & 99\%  & 4\% \\
28     & $^{56}{\rm Ni}$   & 6.075\,d    & Ni & [21-221]  & 21\% & 100\% & 4\%\\
        & $^{59}{\rm Ni}$   &\!\!\!\!81.82\,kyr  & &  & 21\% & 65\%  & 1\% \\
\hline
\end{tabular}
\end{table}

\subsection{EC-decay in IS data}
Only two human-made instruments ventured outside the Solar cavity, namely Voyager\,1 \citep{2013Sci...341..150S} and Voyager\,2 \citep{2019NatAs...3.1019R}, both launched in 1977. \citet{2016ApJ...831...18C} published unmodulated elemental fluxes for H up to Ni (but not Co) from Voyager\,1 in the so-called VLIS (very local interstellar medium). We show in Table~\ref{tab:CRDATA_IS_fluxes} the corresponding EC-unstable isotopes (first 3 columns), along with the energy range (few tens to two hundreds MeV/n) and best precision ($\sim 20\%$) reached by the Voyager instruments on the associated fluxes (next 3 columns). The next-to-last column is our calculated decayed fraction, calculated at 20\,MeV/n (the lowest Voyager data point). The latter is multiplied by the CR isotopic fractions (see Table~\ref{tab:CRDATA_TOA_ISOTOPIC_RATIOS}) to obtain the expected decayed fraction in the elemental flux (last column). To be detectable, this fraction must be larger than the instrument precision: only V and Cr have significant decayed fractions, and V is the best candidate (to be precise, V first data point is at 129\,MeV/n, for which the decayed fraction of $^{49}$V is 78\% and that of V is 31\%). On closer inspection, the $\sim20\%$ CR data precision (on $Z\leq26$ fluxes) automatically excludes all species with isotopic fractions smaller than this number. Among, the remaining ones, $^{49}$V and $^{51}$Cr are the most favourable (short-lived nuclei with $\sim 100\%$ decayed fractions), while $^{7}$Be is too light to attach electrons (and decay), and $^{53}$Mn is too long-lived to significantly decay. Note that $^{57}$Co ought to be a good candidate (short-lived with an isotopic fraction of 0.32), but Co is not available in Voyager data, unfortunately. We stress that the CR isotopic fraction data have a dispersion (not shown) much larger than the weighted error reported in~Table~\ref{tab:CRDATA_TOA_ISOTOPIC_RATIOS}, which could change the detectability of EC-decay in IS V and Cr fluxes.

\begin{table*}[t]
\small
\centering
\caption{Energy range, variance-weighted mean, and best achieved precision on existing CR TOA data ratios involving EC-unstable species: parent-to-daughter ratios \citep{2005PhDT.........6S} and isotopic abundances (same references and weighted-mean already reported in Table~\ref{tab:CRDATA_TOA_ISOTOPIC_RATIOS}).} \label{tab:CRDATA_TOA_ISOTOPES}
\begin{tabular}{rlr@{\hskip 15mm}lccr@{\hskip 15mm}lcr}
\hline\hline
\multicolumn{3}{c}{\hspace{-1.cm}EC-unstable species} & \multicolumn{4}{c}{\hspace{-1.2cm}Parent/daughter (TOA data)} & \multicolumn{3}{c}{Isotopic ratio (TOA data)} \\
$Z$ & Name & $t_{1/2}$ & Qty & $E_{\rm k/n}$\,[GeV/n] &Mean& \!Prec.\!& Qty & $E_{\rm k/n}$\,[GeV/n] & Prec. \\
\hline
4      & $^{7}{\rm Be}$  & 53.22 d   & $^{7}{\rm Be}$/$^{7}{\rm Li}$   &     -       &   -  & -  & $^{7}{\rm Be}$/Be &[0.08-1.42]  & 2\%\\
18     & $^{37}{\rm Ar}$ & 35.011 d  & $^{37}{\rm Ar}$/$^{37}{\rm Cl}$ &[0.12-0.35]  & 1.40  & 9\% & $^{37}{\rm Ar}$/Ar &[0.49-0.57]  & 6\%\\
20     & $^{41}{\rm Ca}$ & 99.4 kyr  & $^{41}{\rm Ca}$/$^{41}{\rm K}$  &[0.12-0.37]  & 0.855 & 10\% & $^{41}{\rm Ca}$/Ca &[0.58-0.65]  & 6\%\\
22     & $^{44}{\rm Ti}$ & 59.1 yr   & $^{44}{\rm Ti}$/$^{44}{\rm Ca}$ &[0.13-0.39]  & 0.048 & 14\% & $^{44}{\rm Ti}$/Ti &[0.32-0.69]  & 27\%\\
23     & $^{49}{\rm V}$  & 330 d     & $^{49}{\rm V}$/$^{49}{\rm Ti}$  &[0.14-0.41]  & 3.10  & 8\% & $^{49}{\rm V}$/V &[0.32-0.69]  & 19\%\\
24     & $^{51}{\rm Cr}$ & 27.7015 d & $^{51}{\rm Cr}$/$^{51}{\rm V}$  &[0.14-0.42]  & 2.52  & 7\% & $^{51}{\rm Cr}$/Cr &[0.32-0.72]  & 10\%\\
25     & $^{53}{\rm Mn}$ & 3.7 Myr   & $^{53}{\rm Mn}$/$^{53}{\rm Cr}$ &[0.14-0.42]  & 3.60  & 9\% & $^{53}{\rm Mn}$/Mn &[0.32-0.74]  & 10\%\\
26     & $^{55}{\rm Fe}$ & 2.7562 yr & $^{55}{\rm Fe}$/$^{55}{\rm Mn}$ &[0.14-0.43]  & 1.52  & 11\% & $^{55}{\rm Fe}$/Fe &[0.62-0.94]  & 10\%\\
28     & $^{56}{\rm Ni}$ & 6.075 d   & $^{56}{\rm Ni}$/$^{56}{\rm Fe}$ &     -       &   -   & -  & $^{56}{\rm Ni}$/Ni &[0.65-0.99]  & 74\%\\
27     & $^{57}{\rm Co}$ & 271.811 d & $^{57}{\rm Co}$/$^{57}{\rm Fe}$ &[0.15-0.44]  & 0.062 & 16\% & $^{57}{\rm Co}$/Co &[0.32-0.32]  & 30\%\\
28     & $^{59}{\rm Ni}$ & 81.82 kyr & $^{59}{\rm Ni}$/$^{59}{\rm Co}$ &     -       &   -   & -  & $^{59}{\rm Ni}$/Ni &[0.10-0.50]  & 70\%\\
31     & $^{67}{\rm Ga}$ & 3.2617 d  & $^{67}{\rm Ga}$/$^{67}{\rm Zn}$ &     -       &   -   & -  & $^{67}{\rm Ga}$/Ga &[0.13-0.70]  & 58\%\\
32     & $^{71}{\rm Ge}$ & 11.43 d   & $^{71}{\rm Ge}$/$^{71}{\rm Ga}$ &     -       &   -   & -  & $^{71}{\rm Ge}$/Ge &[0.13-0.70]  & 64\%\\
33     & $^{73}{\rm As}$ & 80.30 d   & $^{73}{\rm As}$/$^{73}{\rm Ge}$ &     -       &   -  & -  & $^{73}{\rm As}$/As &[0.13-0.70]  & 65\%\\
\hline
\end{tabular}
\end{table*}

\subsection{From IS to TOA decaying fractions}
All previous calculations were performed on IS energies. When we consider TOA fluxes, the generic source term no longer factors out in the decayed fraction. Indeed, Solar modulated fluxes critically depend on the spectral shape of the IS fluxes. We can go around this issue as follows. 
\begin{enumerate}
    \item We assume that the range of possible IS flux shapes is encompassed by the shapes of He to Ni IS fluxes, as parametrised in \citet{2019ApJ...887..132S,2025ApJ...988..262S}. We denote the corresponding fluxes by $\psi^{\rm IS}_{\rm Chen}$.
    
    \item We model the decayed fraction by a generic sigmoid function
    \[
    \Sigma = (1-f_{\rm min}) \times \left[1 - \frac{1}{1 + \exp\left[-k \log(E_{\rm k/n}) - \log(E_0))\right]}\right]\,.
    \]
    We checked that the latter reproduces the variety of decayed fraction shapes. The parameters of the sigmoid control the maximal decayed fraction at low energy, $f_{\rm min}$, and the width, rise, and energy position at which the sigmoid goes to zero at a few GeV/n.

    \item We calculate the TOA decaying fraction by modulating $\psi^{\rm IS}_{\rm no~decay}=\psi^{\rm IS}_{\rm Chen}$ and $\psi^{\rm IS}_{\rm EC~decay}=(1-\Sigma)\,\psi^{\rm IS}_{\rm no~decay}$ (TOA flux without and with decay, respectively), and then forming their relative difference; this gives $f^{\rm TOA}$.
\end{enumerate}
We checked (not shown) that $f^{\rm TOA}$ is insensitive to the IS spectral shape (i.e., whether it is a primary or secondary species) and to the sigmoid parameters (i.e., whether it has a mild, strong, steep or flat decrease). All together, the TOA decayed fraction matches the IS one with the energy shifted by the Solar modulation parameter (see Eq.~\ref{eq:def_ETOA_EIS}):
\[
f^{\rm TOA}(E_{\rm k/n}^{\rm TOA}) = f^{\rm IS}(E_{\rm k/n}^{\rm IS}=E_{\rm k/n}^{\rm TOA}+\phi_{\rm FF}\cdot  Z/A)\;.
\]
The above relation implies that, for the same energy $f^{\rm TOA}\!<\!f^{\rm IS}$. It also means that $f^{\rm TOA}$ is larger for periods of low Solar activity than for high Solar activity ones.

\subsection{EC-decay in TOA isotopic data}

TOA CR data have been obtained for different levels of Solar activity. In this section, for simplicity and to be conservative, we model the decayed fraction for a low level of Solar activity, $\phi_{\rm FF}=500$\,MV, corresponding to an energy shift of $\sim 250$\,MeV/n, and thus a TOA decayed fraction as large as possible.

\paragraph{Data status}
As already said, isotopic separation is hard to achieve in CR experiments. Furthermore, fluxes are more difficult to measure than ratios: the former requires an excellent knowledge of the detector acceptance, whereas the latter does not (and also cancels out many systematics). As a result, almost all isotopic CR data are ratios, that we gather in Table~\ref{tab:CRDATA_TOA_ISOTOPES}.
First, we have isotopic ratios (4th to 6th columns), formed from the EC-unstable parent to their daughter isotope. These ratios maximise the decay impact (decayed parent in the numerator and EC-fed daughter in the denominator), but are not always provided by the experiments. The largest and most precise dataset is that of ACE-CRIS \citep{2005PhDT.........6S}, with a precision varying from 7\% to 16\%.
Second, we also have isotopic ratios (last 3 columns), which are more systematically provided. We clearly see the pattern of decreasing precision with the isotope mass, as both their abundance and isotopic separation ($\Delta A/A$) decrease with $A$, making them harder to measure\footnote{Actually,  the data precision is a non-trivial balance between the instrument mass resolution, the isotopic abundance of the EC-unstable species, and the relative abundance of the latter with respect to the neighbour isotopes. The trend seems to be dominated by the mass resolution and abundance of the EC-unstable under scrutiny, with a poorer precision for heavier isotopes.}. These ratios have been measured by numerous experiments, the most precise one being ACE-CRIS'~\citep{2001ICRC....5.1751C, 2009ApJ...695..666O, 2022ApJ...936...13B}.

\begin{figure}
\centering
\includegraphics[width=0.49\textwidth]{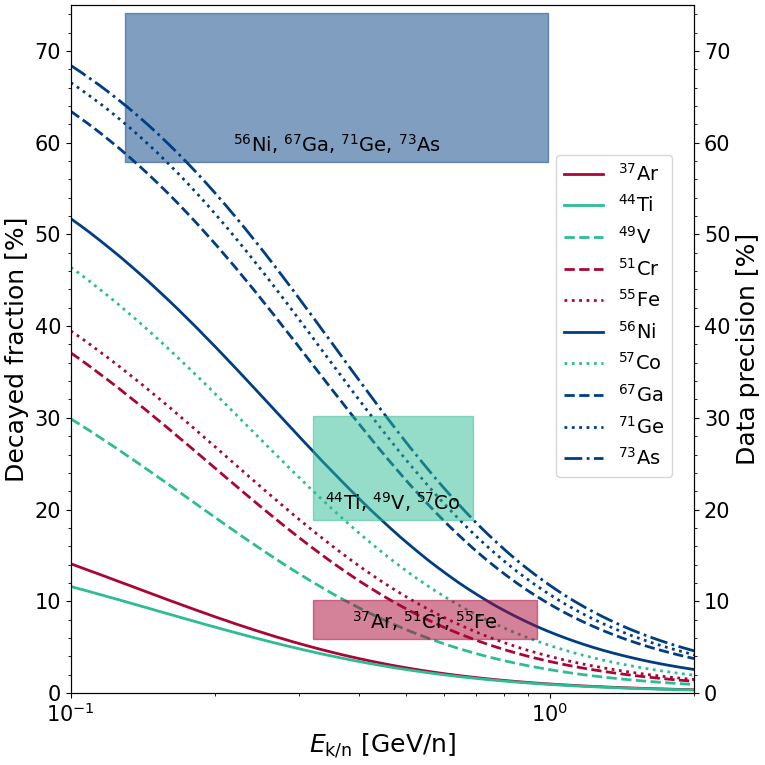}
\caption{Decaying fraction of EC-unstable isotopes (lines) at $\phi_{\rm FF}=500$\,MV as a function of TOA energies, compared to the TOA CR data precision (rectangular boxes). We do not show $^{7}$Be, $^{41}$Ca, $^{53}$Mn, and $^{59}$Ni, whose decayed fraction is below the percent level, and thus largely below the data precision. In order not to overcrowd the figure, we also grouped species whose CR data are of similar precision (from Table~\ref{tab:CRDATA_TOA_ISOTOPIC_RATIOS}) in the same rectangles.}
\label{fig:decfrac_TOA}
\end{figure}

\paragraph{EC-decay detectability in parent nuclei}

We show in Fig.~\ref{fig:decfrac_TOA} the comparison between the decayed fraction computed in the 1D-DM model (lines) and the best data precision (filled rectangles). As before, we say that EC-decay is detectable if the decaying fraction of an isotope is larger than its data relative precision.
First, light EC-unstable species are very rare, with only $^7{\rm Be}$ for $Z < 15$. Even though the isotopic ratio of the latter is measured with the greatest precision ($\sim 2\%$), its decay fraction is less than 1\% (owing to its long attachment time). Second, for $Z>15$, we have different behaviours depending on the isotopes half-life. The decay fraction for intermediate-lived isotopes is strongly dependent on the decay time: $^{41}$Ca ($t_{1/2}=99.4$\,kyr), $^{53}$Mn ($t_{1/2}=3.7$\,Myr) and  $^{59}$Ni ($t_{1/2}=81.82$\,kyr) have a decay fraction at the percent level (not shown) above a few hundreds of MeV/n, while $^{44}$Ti ($t_{1/2}=59.1$\,yr) reaches 10\% (solid cyan line). 
For short-lived isotopes ($t_{1/2}\lesssim$\,yr), the decaying fraction strongly depends on the atomic number, with the heaviest species ($^{67}$Ga, $^{71}$Ge and $^{73}$As) showing the highest decayed fraction.
\begin{table*}[t]
\small
\centering
\caption{Minimal energy and best precision achieved on elemental CR TOA data, either for the fluxes ($4^{\rm th}$ to $6^{\rm th}$ columns) or the parent-to-daughter elemental flux ratios (last 3 columns). For $Z\leq30$, the most precise data are from ACE-CRIS \citep{2013ApJ...770..117L} and AMS-02 \citep{2021PhRvL.126d1104A, 2021PhR...894....1A}; note that AMS-02 data for $22<Z<25$ and $Z>26$ have not been released yet. For Ga/Zn, Ge/Ga and As/Ge, only a few data points exist from TIGER \citep{2009ApJ...697.2083R}, Super-TIGER \citep{2016ApJ...831..148M}, ACE-CRIS \citep{2022ApJ...936...13B} and CALET \citep{2025ApJ...988..148A}. More CR data exist for heavier elemental ratios, but their isotopic content is not measured, so that the decayed fraction cannot be estimated. Furthermore, we stress that the best precision reported in this table is not necessarily obtained at the lowest energy reported.\label{tab:CRDATA_TOA_ELEMENTS}}
\begin{tabular}{rlr@{\hskip 15mm}lcc@{\hskip 15mm}lcr}
\hline\hline
 \multicolumn{3}{c}{\hspace{-1cm}EC-unstable species} & \multicolumn{3}{c}{\hspace{-1.4cm}Elemental flux (TOA data)} & \multicolumn{3}{c}{\hspace{-2mm}Elemental ratio (TOA data)} \\
 $Z$ & Name &  $t_{1/2}$ & Qty&\!\!\!\!$E_{\rm k/n}$\,[GeV/n]\!\!\!\!& rel.err & Qty &\!\!\!\!$E_{\rm k/n}$\,[GeV/n]\!\!\!\!& rel.err \\
\hline
4      & $^{7}{\rm Be}$    & 53.22 d   & Be &0.06 & 2\% & Be/Li & 0.08 & 6\%\\
18     & $^{37}{\rm Ar}$   & 35.011 d  & Ar &0.12 & 3\% & Ar/Cl & 0.12 & 6\%\\
20     & $^{41}{\rm Ca}$   & 99.4 kyr  & Ca &0.14 & 3\% & Ca/K & 0.14 & 5\%\\
22     & $^{44}{\rm Ti}$   & 59.1 yr   & Ti &0.14 & 3\% & Ti/Ca &  0.14 & 5\% \\
23     & $^{49}{\rm V}$    & 330 d     & V &0.14 & 4\% & V/Ti & 0.14 & 6\%\\
24     & $^{51}{\rm Cr}$   & 27.7015 d & Cr &0.14 & 3\% & Cr/V & 0.14 & 6\%\\
25     & $^{53}{\rm Mn}$   & 3.7 Myr   & Mn &0.14 & 4\% & Mn/Cr & 0.14 & 5\%\\
26     & $^{55}{\rm Fe}$   & 2.7562 yr & Fe &0.17 & 2\% & Fe/Mn & 0.17 & 5\%\\
28     & $^{56}{\rm Ni}$   & 6.075 d   & Ni &0.17 & 5\% & Ni/Fe &  0.17 & 5\% \\
27     & $^{57}{\rm Co}$   & 271.811 d & Co &0.17 & 5\% & Co/Fe & 0.17 & 5\%\\
28     & $^{59}{\rm Ni}$   & 81.82 kyr & Ni &0.17 & 5\% & Ni/Co & 0.17 & 5\%\\
31     & $^{67}{\rm Ga}$   & 3.2617 d  & Ga & - & -  & Ga/Zn & 0.30 & 12\%\\
32     & $^{71}{\rm Ge}$   & 11.43 d   & Ge & - & -  & Ge/Ga & 0.30 & 13\%\\
33     & $^{73}{\rm As}$   & 80.30 d   & As & - & -  & As/Ge & 0.30 & 21\%\\
\hline
\end{tabular}
\end{table*}

All in all, the detectability of EC-decay is possible if the curves are below the corresponding boxes (of same colour). This only clearly happens for $^{51}$Cr and $^{55}$Fe (dashed and dotted red lines), and otherwise, we barely meet the condition for $^{57}$Co (thin cyan line), and 
$^{67}$Ga, $^{71}$Ge and $^{73}$As (dashed, dotted, and dash-dotted blue lines). So it seems that we are at the verge of being able to detect EC-decay in TOA CR data. However, this must be mitigated by the fact that EC-decay is overestimated in our modelling, as we neglected energy losses. And this is without considering the impact of nuclear production \citep{1993ApJ...403..644C, 1996ApJ...470.1218W, 1998NewAR..42..277W} and inelastic cross-section uncertainties \citep{2025arXiv250316173M}, possibly larger than the decayed fraction itself.

\paragraph{EC-decay detectability in daughter nuclei}
EC decay could also be sought in the daughter of the EC-unstable species. The weighted-average mean of the parent-to-daughter ratio (denoted $r$ in this paragraph), reported in Table~\ref{tab:CRDATA_TOA_ISOTOPES} (6th column), gives an indication whether this would be more favourable than looking at the parent. Indeed, the daughter EC-fed fraction, $f_d$, is  simply linked to the EC-decayed fraction of the parent, $f_p$, by $f_d=r\times f_p$. For ratios with $r<1$ (e.g., $^{41}$Ca/$^{41}$K), EC decay will be even harder to see in the daughter than in the parent EC-unstable species. This is the reverse for ratios with $r>1$, and five ratios meet this criterion in Table~\ref{tab:CRDATA_TOA_ISOTOPES}. We provide in Table~\ref{tab:CRDATA_TOA_ISOTOPIC_DAUGHTER} their $f_p$ and $f_d$ values: the already spotted $^{49}$V/$^{49}$Ti and $^{51}$Cr/$^{51}$V ratios become even more favourable than before, thanks to the increased EC-decay impact in the daughter (in the denominator). In addition, $^{55}$Fe/$^{55}$Mn now passes the detectability criterion, with $f_d$ larger than the CR data precision ({\em rel. err.} column in the Table). These trends would need to be confirmed by a complete analysis involving the full nuclear network. Indeed, Ti, V, and Mn isotopes are of secondary origin, so that they are maximally sensitive to the cross-section uncertainties mentioned earlier.

\begin{table}[t]
\small
\centering
\caption{Impact of EC-decay in parent-to-daughter ratios, for cases in which the detectability is larger in the daughter than in the parent. The first three columns (directly reported from Table~\ref{tab:CRDATA_TOA_ISOTOPES}) correspond to the CR TOA data ratio, weighted-average (denoted $r$), and best data precision. The next-to-last column reports the decayed fraction $f_p$ in the parent, as read off Fig.~\ref{fig:decfrac_TOA}, at the most favourable energy $\sim150$\,MeV/n where data exist. The last column reports the decayed fraction fed to the daughter, $f_d = r\times f_p$, at this same most favourable energy}.
\label{tab:CRDATA_TOA_ISOTOPIC_DAUGHTER}
\begin{tabular}{rcccl}
\hline\hline
Ratio & $r$ & rel.err. & $f_p$ & $f_d$\\
\hline
$^{37}$Ar/$^{37}$Cl &  1.40 & 9\% & 6\%    & 8.4\% \\
$^{49}$V/$^{49}$Ti  &  3.10 & 8\% & 12\%   & 37.2\%\\
$^{51}$Cr/$^{51}$V  &  2.52 & 7\% & 16\%   & 40.3\%\\
$^{53}$Mn/$^{53}$Cr &  3.60 & 9\% & $<1\%$ & $<3.6\%$\\
$^{55}$Fe/$^{55}$Mn &  1.52 & 11\%& 20\%   & 30.4\%\\
\hline
\end{tabular}
\end{table}

\subsection{EC decay in TOA elemental data}

It is also interesting to study our results in the context of available data on elemental fluxes, which are generally easier to measure (they bypass the challenges of isotopic separation), and therefore can be obtained with higher precision. 

\paragraph{Data status}
Elemental fluxes have been measured by a huge number of experiments. But at low energy and for $Z\leq30$, the most precise data have a few percent precision, and are obtained down to $\simeq 100$\,MeV/n from ACE-CRIS \citep{2013ApJ...770..117L} and AMS-02\footnote{To be precise, AMS-02 data are in rigidity unit, and converting them in kinetic energy per nucleon introduces additional systematics at the few percent level \citep{2019A&A...627A.158D}, because the isotopic content is not measured above a few GeV/n.} with a minimum energy of $\simeq 500$\,MeV/n~\citep{2021PhRvL.126d1104A, 2021PhR...894....1A}. A summary view of the relevant data is gathered in Table~\ref{tab:CRDATA_TOA_ELEMENTS}, for the elemental fluxes (4th to 6th columns), but also for ratios of parent/daughter elemental fluxes (last 3 columns) -- the latter ratios contain the EC-unstable isotope in their numerator and its daughter in their denominator, maximising the EC-decay impact. Only ratios have been measured for $Z>30$, by TIGER \citep{2009ApJ...697.2083R}, Super-TIGER \citep{2016ApJ...831..148M}, ACE-CRIS \citep{2022ApJ...936...13B} and CALET \citep{2025ApJ...988..148A}. These experiments provide data up to $Z\simeq40$, but that cannot be exploited here: the isotopic content of EC-unstable species is necessary to obtain the elemental decaying fraction, and this content has only been measured\footnote{In principle, we could use the predicted isotopic content instead, but we do not have it in our very simplified approach.} for $Z=31-33$ \citep{2022ApJ...936...13B}.

\paragraph{EC-decay detectability}
\begin{figure}
\centering
\includegraphics[width=0.49\textwidth]{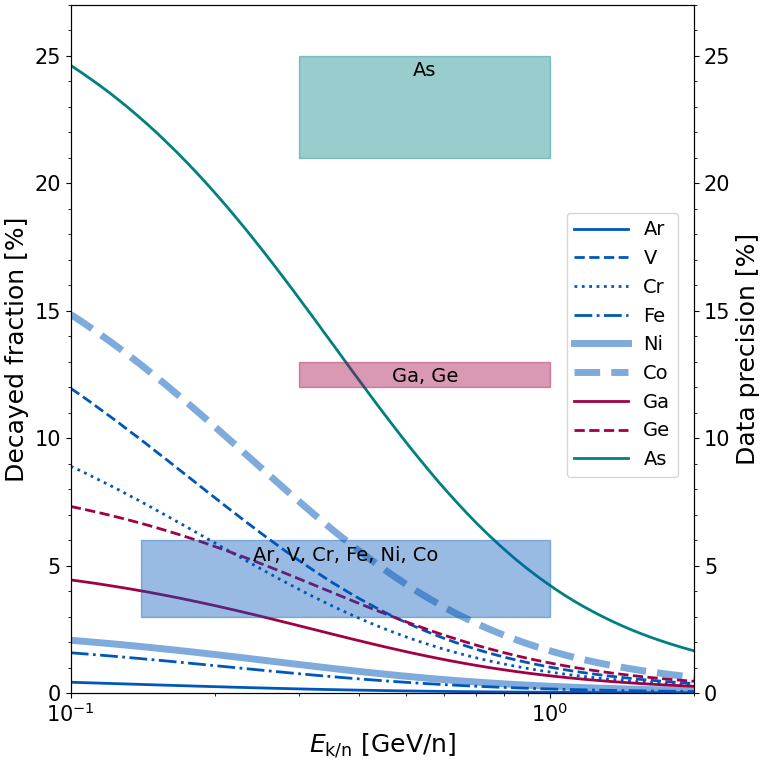}
\caption{Same as in Fig.~\ref{fig:decfrac_TOA}, but for the decayed fraction in TOA elemental fluxes (compared to TOA CR data). We do not show Be, Ca, Ti, Mn, and Ni, whose decayed fraction is below the sub-percent level, and thus largely below the data precision.}
\label{fig:decfrac_TOA_elements}
\end{figure}

As discussed in the previous section, the TOA decayed fraction of intermediate-lived isotopes $^{7}$Be, $^{41}$Ca, $^{53}$Mn, and $^{59}$Ni are at the percent level, so when diluted by their isotopic content (0.57, 0.07, 0.44 and 0.02, respectively, see Table~\ref{tab:CRDATA_TOA_ISOTOPIC_RATIOS}), this fraction is sub-percent in the elemental TOA fluxes (Be, Ca, Mn, and Ni, respectively). Even the decayed fraction in Ti is sub-percent, owing to its small isotopic fraction: the decayed fraction of $^{44}$Ti is 10\%, diluted by its 0.01 isotopic fraction. Only short-lived EC-unstable isotopes provide $\gtrsim 1\%$ decayed fractions. We show, in Fig.~\ref{fig:decfrac_TOA_elements}, their expected decayed fractions (lines), compared to the CR data precision (rectangular boxes). For $Z\leq30$, only V, Cr, and Co (thin-dashed, dotted, and thick-dashed blue lines, respectively) have data precision at the level or better than the expected decayed fraction, Co being the most favourable case. Actually, forthcoming AMS-02 data on these species will start at $\gtrsim 500$\,MeV/n with a $\gtrsim 6\%$ (uncertainties dominated by systematics on the acceptance), so ACE-CRIS data are more favourable to search for this decay. For $Z>30$, neither Ga and Ge (red lines), nor As (cyan line) meet this requirement, and by a factor of two or more.

\section{Conclusions}
\label{sec:Conclusions}

We have revisited the impact of EC decay on isotopic and elemental fluxes, assessing the detectability of this decay in current IS and TOA CR data.

First, on the modelling side, we obtained analytical solutions accounting for multiple ionic states of CRs during propagation. We showed that the widely used 2-level approximation (i.e., fully ionised CR or single electron attached) differs from the full calculation (by $\lesssim 50\%$) only for intermediate-lived nuclei $t_{1/2}\sim {\cal O}(10\,{\rm kyr})$ with $Z>40$ and at IS energies $\lesssim 100$\,MeV/n. There are actually very few EC-unstable species in this decay time and charge range. Furthermore, intermediate-lived nuclei are found to have mild decaying fractions, well below the precision of current (and possibly forthcoming) data. As a result, for all practical purpose, the 2-level approximation is deemed appropriate.

Second, we have made a quantitative study of the impact of attachment and stripping cross-section uncertainties on the decaying fractions. Short-lived species (i.e., $t_{1/2}<{\cal O}({\rm kyr})$) are very sensitive to uncertainties on the attachment cross-sections (timescales $\gtrsim$\,kyr), but insensitive to those on stripping cross-sections. The latter only impacts intermediate-lived nuclei (also impacted by attachment cross-section uncertainties), but only mildly. We stress that the cross-sections used in the literature and in this paper were derived more than forty years ago. While their values are not critical, unless they are off by a factor of a few, updating them and assessing their accuracy and precision would certainly be an improvement.

Third, we have compared the decaying fraction in the still widely used LBM (for $Z>30$ analyses) and in a more realistic diffusion model. The leakage-lifetime approximation (at the heart of the LBM) is known to fail for $\beta$-unstable species, and it also fails for EC-unstable species. Nevertheless, the decaying fraction trend as a function of energy, charge, and decay times, are qualitatively similar in both models. Differences on the fluxes can be as high as a factor of a few, but in a parameter space where very few EC-unstable species actually lie.

Last, we have compared the decayed fraction to both IS and TOA current data precision. While IS data are more favourable in terms of energy, only elemental fluxes -- in which the isotope decay is diluted by the presence of the other stable isotopes -- have been measured. Low-energy TOA data are more numerous and benefit from the latest generation of high-precision experiments (mainly ACE-CRIS for isotopes and fluxes, and  AMS-02 for fluxes). All in all, only $^{49}$V, $^{51}$Cr and $^{57}$Co meet the EC-decay detectability condition -- i.e., the predicted decayed fraction is larger than the current data precision --, in both isotopic TOA data and elemental IS or TOA fluxes. Actually, the daughters of $^{49}$V, $^{51}$Cr and $^{55}$Fe, fed by their parent decays, are even more significantly impacted (than their parents): parent-to-daughter ratios (of these species) are the best hope for a detection at this stage. We stress that all our calculations were performed neglecting energy losses, so that the decayed fractions are probably slightly overestimated.
This means that the results obtained are conservative with respect to the non-detectability cases, but detectable ones need to be confirmed by a more rigorous calculation.

In summary, while the impact of EC decay remains difficult to confirm in current CR data, the isotopes of V, Cr, and Co are the most promising candidates to do so. Several aspects of our calculation require refinement, though: first,
the model must incorporate energy losses and solve the transport equation with the complete network of nuclear production cross-sections; second,
uncertainties on nuclear cross-sections for $Z>30$ must be quantified, as they are known to bring large uncertainties in the modelling, especially for pure secondary species.
In any case, the presence of many short-lived EC-unstable isotopes (with $Z>30$) that should be fully decayed, combined with the prospect of measurements in the near future (e.g., from ISS-TIGER), provides a strong motivation to investigate further this topic.

\section*{Acknowledgments}
We thank the referee for their insightful comments.
This work is part of the project ``Cosmic ray antideuterons as a probe for new physics'' with project number OCENW.KLEIN.387 (Budget Number 11680) of the research programme Grant Open Competition Domain Science, which is financed by the Dutch Research Council (NWO). The work of D.~Maurin was supported by the INTERCOS project funded by IN2P3.

\begin{appendix}
\section{Attachment and stripping cross-sections}
\label{app:sig_att_str}

Electron attachment and stripping cross-sections are essential to properly describe the transport of EC-unstable species. We start below with the formulae for a single electron attachment (hydrogen-like ions with only two possible charged states) and discuss the corrections needed to extend them to a multilevel model. We recall that these cross-sections are summed over the relevant ISM component densities, see Eq.~\eqref{eq:gamma_att_str}.

Electron attachment is denoted electron capture in the literature \citep{2007PhR...439....1E}, but to avoid confusion with the EC-decay process, we keep the former denomination throughout this paper. We use (and report below) the cross-section parametrisations of \citet{1984ApJS...56..369L}, derived in the context of CR studies. The value and precision of these cross-sections should probably be revisited, in the light of the progress made in EC in ion-atom collisions; see \citet{2023PhyU...66.1177T} for a recent, comprehensive, though broader review. But here, we keep these parametrisations for comparison purpose, as they have been used in all previous CR studies. We nevertheless provide a quantitative estimate of the impact (on the propagated CR fluxes) of these cross-section uncertainties in \ref{sec:impact_att_str_uncertainty}, to assess how critical they are.

\subsection{Attachment}
\label{app:sig_att}
The capture of an electron, from a neutral ISM target T, by a CR projectile P, is a combination of radiative attachment (ra) and non-radiative attachment (nra),
\begin{equation}
    \sigma^{\rm att}_{\rm P+T} = \sigma^{\rm ra}_{\rm P+T} + \sigma^{\rm nra}_{\rm P+T}\,,
\end{equation}
with
\begin{eqnarray}
{\rm P}^{q+} + {\rm T} &\rightarrow& {\rm P}^{(q-1)+} + {\rm T}^+ + \gamma\,
\quad {\rm (ra)}\,,\nonumber\\
{\rm P}^{q+} + {\rm T} &\rightarrow& {\rm P}^{(q-1)+} + {\rm T}^+ \quad\quad\;\,{\rm (nra)}\,. \nonumber
\end{eqnarray}
In the above reactions, $q+$ is the initial charge of the projectile, with $q+=Z_{\rm P}$ if the projectile is completely ionised. In the non-radiative process, the excess momentum of the electron is carried out by the photon (3-body problem). In the non-radiative one, the momentum is conserved through the recoil of the target T (from which the electron was captured).
These two processes have similar energy-dependences in the asymptotic relativistic regime, with $\sigma^{\rm ra}_{\rm P+T} \propto Z_{\rm P}^5 Z_{\rm T}/E$ and $\sigma^{\rm nra}_{\rm P+T} \propto Z_{\rm P}^5 Z_{\rm T}^5/E$, but strongly differ in the highly non-relativistic one, with $\sigma^{\rm ra}_{\rm P+T} \propto \beta^{-5}$ and $\sigma^{\rm nra}_{\rm P+T} \propto \beta^{-12}$ -- for a thorough discussion, see \citet{2007PhR...439....1E}

\paragraph{Extension of the 1s--1s attachment formula}
The starting point is to consider the attachment of an electron from the K-shell (or subshell 1s) of a single-electron target atom, to the K-shell of a fully stripped projectile, $\sigma^{\rm att}_{\rm P(1s) \,+\, T(1s)}$, whose formula is detailed in the next paragraph.
Accounting for the attachment in all other projectile shells -- which falls off as $1/n^3$ for the $n$-th shell~\citep{1979PhDT........67C} --, we have
\begin{equation}
  \sigma^{\rm att}_{\rm P\,+\,T(1s)} = \sum_{n=1}^{\infty} \frac{1}{n^3} \,\sigma^{\rm att\;(1s)}_{\rm P(1s)\,+\,T(1s)} = 1.202 \, \sigma^{\rm att\;(1s)}_{\rm P(1s)\,+\,T(1s)}\,.
\end{equation}
In the case of targets with more than one K-shell electron (or electrons also in other shells), one has to take into account the extra number of electrons available, and also the partial screening of the nuclear potential due to the presence of these extra electrons. This screening leads to an effective charge of the target~\citep{1963JChPh..38.2686C,1967JChPh..47.1300C,2017ADNDT.117..439G}, with $Z_{\rm T}^{\rm eff} \approx Z_{\rm T} - 0.3$ a reasonable approximation for practical calculations. In the context of H and He targets in the ISM, the relevant formulae to use are
\begin{eqnarray}
\sigma^{\rm att}_{\rm P+HI} &=& 1.202\,\sigma^{\rm att}_{\rm P(1s)\,+\,T(1s)}(Z_{\rm T}=1)\,; \\
\sigma^{\rm att}_{\rm P+He} &=& 2\times 1.202\,\sigma^{\rm att}_{\rm P(1s)\,+\,T(1s)}(Z_{\rm T}^{\rm eff}=1.7)\,.
\end{eqnarray}

\paragraph{1s--1s radiative attachment} The cross-section $\sigma^{\rm ra}_{\rm P(1s)\,+\,T(1s)}$ can be calculated from the photoionisation cross-section \citep{1979PhDT........67C}, 
\[
{\rm P}^{q+} + \gamma \rightarrow {\rm P}^{(q+1)+} + e^-\,,
\]
using the method of detailed balance, assuming there are two electrons available in the K-shell. As a consequence, this cross-section has to be divided by two to get the one for the attachment of a single electron. For a fully stripped projectile, this cross-section reads \citep{1979PhDT........67C, 1984ApJS...56..369L}:
\begin{eqnarray}
\label{eq:rad_att}
     \sigma^{\rm ra}_{\rm P(1s)\,+\,T(1s)} \!\!&=&\!\! \frac{3}{2} \, \sigma_T \, Z_{\rm P} \, Z_{\rm T} \,\frac{\beta \, \gamma}{(\gamma - 1)^3}\, A^{4+2\xi} \, \exp \left( \frac{-2 A}{\beta \cos A}\right) \nonumber\\
     & \times&\bigg( M(\beta) \; \Big[1 \, + \, R(A)\Big] \, + \pi\, A\, N(\beta)\bigg)\,,
\end{eqnarray}
where $\sigma_T$ is the Thompson cross-section, $Z_{\rm P}$ is the projectile charge, $Z_{\rm T}$ the target charge, $A = \alpha\,Z_{\rm P}$ (with $\alpha=0.00729$ the fine structure constant), and

\begin{eqnarray}
     \xi &=& \left(1\,-\,A^2\right)^{1/2} - 1\,,\nonumber \\
     M(\beta) &=& \frac{4}{3} + \frac{\gamma \, (\gamma -2)}{(\gamma +1)} \, \left[ 1 - \frac{1}{2\beta\gamma^2} \, \ln\left( \frac{1+\beta}{1-\beta}\right) \right]\,, \nonumber \\
    N(\beta) &=& \frac{1}{15\beta^3} \left(-4\gamma + 34 - \frac{63}{\gamma} + \frac{25}{\gamma^2} + \frac{8}{\gamma^3}\right) \nonumber\\
    & & - \frac{(\gamma - 2)}{2\beta^2\gamma(\gamma \!+\!1)}  \ln \left( \frac{1\!+\!\beta}{1\!-\!\beta} \right)\,, \nonumber\\
    R(A) &=& -\exp\left(-8.4\,A^2 + 14\,A - 8.28\right)\,. \nonumber
\end{eqnarray}

\paragraph{1s--1s non-radiative attachment} A non-relativistic quantum mechanical treatment (OBK approximation) allows for computing the probability of transferring an electron from the K-shell of a single-electron target atom to the K-shell of a fully stripped nucleus \citep{osti_6508366}:
\begin{equation}
\sigma^{\rm nra\;(1s-1s)}_{\rm P+T} = \frac{\pi}{5} \,\frac{2^{18} \, a_0^2 \, (Z_{\rm P}\, Z_{\rm T})^5 \, \gamma^2 \, S^8}{\left[ S^2 \!+\! (Z_{\rm P} + Z_{\rm T})^2\right]^5 \, \left[ S^2 + (Z_{\rm P} - Z_{\rm T})^2\right]^5}\,,
\end{equation}
where $S = \beta\gamma / \alpha$ and $a_0 = 5.29 \times 10^{-9}$\,cm is the Bohr radius.

\subsection{Stripping}
\label{app:sig_str}
Stripping processes can be schematically represented as follows:
\begin{eqnarray}
{\rm P}^{q+} + {\rm T} &\rightarrow& {\rm P}^{(q+1)+} + {\rm T} + e^-\,.\nonumber\\
{\rm P}^{q+} + {\rm T}^{p+} &\rightarrow& {\rm P}^{(q+1)+} + {\rm T}^{(p-1)+}\,. \nonumber
\end{eqnarray}
The cross-section for this second process is almost negligible~\citep{1979PhDT........67C}, and the first one is
calculated from Mott and Massey's relativistic ionisation cross-section ~\citep{osti_6508366}. The loss of a K-shell electron by a hydrogenic atom is given by
\begin{equation}
\sigma^{\rm str}_{\rm P+T} = 4 \, \pi a_0^2 \, \left( \frac{\alpha}{Z_{\rm P}\beta} \right)^2 \, (Z_{\rm T}^2 + Z_{\rm T}) \, C_1 \, \left( \ln \frac{4\beta^2 \gamma^2}{C_2\alpha^2 Z_{\rm P}^2} - \beta^2\right)
\end{equation}
where $C_1 = 0.285$ and $C_2= 0.048$ are constants for capture from K-shell.
The $Z_{\rm T}$ term arises from a sum over the contributions from all individual electrons in the target atom, while the $Z_{\rm T}^2$ term accounts for the ionisation by the nucleus. The final stripping cross-section formula is the weighted sum of the cross-sections for the different ISM target abundances. We consider here the contributions of H and He to the ISM target material, with with their number fractions being 0.9 and 0.1, respectively.

\subsection{Extension for the multi-level model}\label{sec:sigma_ext}
All the formulas presented in the previous sections consider just two ionised states for the projectile, with effective charge $q_{\rm P}=Z_{\rm P}$ and $q_{\rm P}=Z_{\rm P}-1$. Indeed, they were specifically derived for attachment of an electron to a fully ionised nucleus or stripping from a hydrogenic atom. 
In our analysis, we generalise the transport equation to a multi-level model (see~\ref{app:sol_multi-level}), 
tracking the evolution of the CR ions up to any level of ionisation state $q_{\rm P}\in \{Z_{\rm P}, (Z_{\rm P}-1), (Z_{\rm P}-2), \dots\}$. This requires the knowledge of the attachment and stripping cross-sections for partly ionised ions. The semi-empirical Schlachter formula proposes $\sigma^{\rm att}\propto q_{\rm P}^{3.9}$  \citep{1983PhRvA..27.3372S}, with an accuracy $\lesssim 2$ \citep{2023PhyU...66.1177T}. As there are large uncertainties on this formula, we keep the description simple and assume the above formulae apply for all ionisation states. This assumption was actually made in the context of the more complex problem of the study of charge states of relativistic heavy ions in matter \citep{1998NIMPB.142..441S}. In any case, we find that only attachment levels up to a few are relevant for most of the species in the CR context (see Sect.~\ref{sec:res_validityapprox}), so this approximation should not be too critical. Another assumption we make is to neglect multielectron capture or stripping -- we assume electrons are attached or stripped one at a time --, whereas mutielectron processes can be significant at almost all energies \citep{2023PhyU...66.1177T}. However, the dependence of the multielectron cross-sections on several parameters (atomic structure of colliding particles, velocity of the incident ion, and other atomic parameters) is still being investigated experimentally and theoretically \citep[e.g.,][]{2025PhRvA.112b2805W}. In the end, for the same reasons as above -- simplicity and the fact that mostly low-level attachments and stripping seem to be relevant for CR --, this assumption should be good enough in the CR context.

\section{Solutions of the transport equation}
\label{app:solutions}

\subsection{Multi-level solution for EC decay}
\label{app:sol_multi-level}
We consider below a single EC-unstable CR species, and we assume a generic source term ${\cal S}$ (of primary or secondary origin), providing the ionised state $N_0$ only. In this study, we neglect energy redistribution terms, and for simplicity, we also discard the convective term in the 1D-DM. The multi-level (up to level $n$) transport equation for the coupled ionised states, for any of the above models, can be written generically (with $j=1,\dots,n-1$)
\begin{equation} \label{eq:multilevel}
    \begin{cases}
        {\cal O} N_0 \!+\! {\cal F}\left[\left(\Gi + \Ga\right) N_0 - \Gs N_1\right]  = {\cal S}\,,\\
		{\cal O} N_j + {\cal F}\left[\left(\Gi \!+\! \Gs \!+\! \Ga \right) N_j - \Ga N_{j-1} - \Gs N_{j+1}\right] \!+\! \GECtilde N_j= 0\,,\\
        {\cal O} N_n \!+\! {\cal F}\left[\left(\Gi + \Gs\right) N_n - \Ga N_{n-1}\right] + \GECtilde N_n = 0\,,
    \end{cases}
\end{equation}
where the operator ${\cal O}$, the factor ${\cal F}$, and the source term ${\cal S}$, for the LBM and the 1D-DM, are given respectively by
\begin{equation} \label{eq: multilevel_factors}
    \begin{aligned}
      \left({\cal O},\, {\cal F},\, {\cal S}\right)_{\rm LBM\;}
        &= \left(\tau_{\rm esc}^{-1},\; 1,\; Q\right)\,,\\
      \left({\cal O},\, {\cal F},\, {\cal S}\right)_{\rm 1D-DM}
        &= \left(-D\,\partial^2_z,\; 2h\,\delta(z),\; 2h\,\delta(z)\,Q\right)\,.
    \end{aligned}
 \end{equation}

The main differences between the 1D-DM and the LBM are the differential transport operator (instead of the escape time) and the attachment, stripping, inelastic cross-sections, and source term tied to the thin disc only instead of the full LB volume. Alternatively, EC decay is possible in the whole confinement volume in both cases.

For the last level at which the multi-level calculation stops, attachment is not enabled, in order to conserve the total number of CRs. Also, we have
\begin{equation}
\label{eq:GEC_halving}
\GECtilde =
    \begin{cases}
     \GEC/2 \quad {\rm for}~i=1\,,\\
     \GEC \quad {\rm otherwise}.
    \end{cases}
\end{equation}
Indeed, the EC half-life measured in the laboratory is for nuclei with two electrons in the K-shell, so that when only one electron is attached (i.e., $i=1$ above), the half-life is doubled and the rate halved. In the above equations, the energy and species (i.e., $A$ and $Z$) dependences are implicit, but are in almost all terms. Following~\ref{app:sig_att_str}, we have $\Ga = \tilde{\Gamma}^{\rm ra} + \Gamma^{\rm nra}$ with
\begin{equation}
\tilde{\Gamma}^{\rm ra} =
    \begin{cases}
     \Gamma^{\rm ra} \quad {\rm for}~n=1\,,\\
     2\Gamma^{\rm ra} \quad {\rm otherwise}.
    \end{cases}
\end{equation}

For the LBM, the above set of equations is directly a coupled set of algebraic equations. For the 1D-DM, this is a set of coupled differential equations, but following the procedure highlighted, e.g., in \cite{2001ApJ...555..585M} -- i.e., solving in the halo with boundary condition for each level, $N_j(z=L)=0$, and integrating through the thin disc ensuring continuity --, we also arrive at a similar set of coupled algebraic equations. The latter can be written in a matrix form, $\underline{\underline{A}}\;\underline{N}\,=\,\underline{S}$,
\begin{equation} 
\label{eq:multilevel_td}
\begin{bmatrix}
\alpha_0 & -\Gs     & 0      & \cdots         & 0        \\
-\Ga     & \alpha_1 & -\Gs   &                & \vdots   \\
0        & \ddots   & \ddots & \ddots         & 0        \\[1mm]
\vdots   &          & -\Ga   &\!\alpha_{n-1}\!& -\Gs     \\[2mm]
0        & \cdots   & 0      & -\Ga           & \alpha_n
\end{bmatrix}
\;
\begin{bmatrix}
N_0 \\[2mm]
N_1 \\[2mm]
\vdots \\[2mm]
N_{n-1} \\[2mm]
N_n
\end{bmatrix}
=
\begin{bmatrix}
Q \\[2mm]
0 \\[2mm]
\vdots \\[2mm]
0 \\[2mm]
0
\end{bmatrix}
\,,
\end{equation}
where the $\alpha$ coefficients are defined to be
\begin{equation}
    \begin{cases}
    \label{eq:def_alpha}
        \alpha_0 = {\cal P} + \Gi + \Ga,\\
        \alpha_j = {\cal V} + \Gi + \Gs + \Ga \quad (j=1,\dots,n-1),\\
        \alpha_n = {\cal V} + \Gi + \Gs,
    \end{cases}
\end{equation}
and with the multiplicative factors ${\cal P}$ and $ {\cal V}$, respectively given by (for the LBM and the 1D-DM)
\begin{equation}
  \label{eq:def_PQ}
    \begin{aligned}
    \left({\cal P},\; {\cal V}\right)_{\rm LBM\;} &=
    \left(\tau_{\rm esc}^{-1},\; \tau_{\rm esc}^{-1}+\GECtilde\right)\,,\\
    \left({\cal P},\; {\cal V}\right)_{\rm 1D-DM} &=
    \left(\frac{D}{hL},\; \sqrt{\frac{D\,\GECtilde}{h^2}} \coth\left(\sqrt{\frac{L^2\GECtilde}{D}}\right)\right)\,.
    \end{aligned}
\end{equation}

Formally, the solution of Eq.~\eqref{eq:multilevel_td} is $\underline{N}\,=\,\underline{\underline{A}}^{-1}\,\underline{S}$. Owing to the simple form of $\underline{S}$, we have $N_i = Q \, (\underline{\underline{A}}^{-1})_{i,0}$. The coefficients of the inverse of a generic tridiagonal matrix can be found, for instance, in \cite{USMANI1994413} or \cite{Huang_1997}. Defining recursively
\begin{equation}
\label{eq:def_theta}
\theta_j = \alpha_j\,\theta_{j+1} - \Ga\Gs\,\theta_{j+2} \quad\quad (j=n-1,n-2,\ldots, 0)
\end{equation}
with $\theta_{n+1}=1$ and $\theta_{n}=\alpha_{n}$, we get $(\underline{\underline{A}}^{-1})_{\,0,0} = \theta_1/\theta_0$
and $(\underline{\underline{A}}^{-1})_{j>0,0} = (\Ga)^j \,(\underline{\underline{A}}^{-1})_{\,0,0}\, (\theta_{j+1}/\theta_1) = (\Ga)^j\, (\theta_{j+1}/\theta_0)$, leading to the solution
\begin{equation}
\label{eq:sol_nlevel}
 N_j = q \, \left(\prod \Ga_1 \dots \Ga_{j-1}\right)\frac{\theta_{j+1}} {\theta_0}\quad(j=0, \dots, n)\,, 
\end{equation}
or alternatively
\begin{equation}
\label{eq:sol_nlevel_bis}
 \begin{cases}
 \displaystyle N_0=Q \, \frac{\theta_1}{\theta_0}\,, \\
 \displaystyle N_j=\Ga_j \left(\frac{\theta_{j+1}}{\theta_j}\right)\times N_{j-1} \quad(j=1, \dots, n)\,,
\end{cases}
\end{equation}

We stress that the dimension of $\theta_j$ is a power-law of time T, $[\theta_j]=$~T$^{j-1-n}$, so that we recover $[N_j]=[Q]$\,T$^{-1}$.

To summarise, the solutions can be computed recursively from Eqs.~\eqref{eq:sol_nlevel} and ~\eqref{eq:def_theta}, with $\alpha_i$ defined in Eq.~\eqref{eq:def_alpha} from the multiplicative factors ${\cal P}$ and $ {\cal V}$ defined in Eq.~\eqref{eq:def_PQ}.
This solution can be extended to a mixed unstable species, with both EC {\em and} $\beta$ decay, by replacing $\GECtilde$ by $(\GECtilde+\gamma^\beta)$ in the $ {\cal V}$ coefficient in Eq.~\eqref{eq:def_PQ}.\footnote{For a pure $\beta$-unstable species, that is no EC and thus $N_0$ only, we recover the DM formula given, for instance, in \citet{2022A&A...667A..25M}.}

\subsection{Formula for a single level ($n=1$)}
Most (if not all) papers in the literature consider a single attachment only, i.e., zero or one electron attached ($n=1$ and $\Ntwolev=N_0+N_1$). In that case, the solution is
\begin{equation}
    \begin{cases}
        \displaystyle N_0 = \frac{Q}{\alpha_0-\Ga\Gs\alpha_1^{-1}}, \\
        \displaystyle N_1 = \frac{\Ga}{\alpha_1}N_0 =\frac{\Ga}{\alpha_1}\times \frac{Q}{\alpha_0-\Ga\Gs\alpha_1^{-1}}.
    \end{cases}
\end{equation}

Substituting $\alpha_i$, ${\cal P}$, and $ {\cal V}$ from Eqs.~\eqref{eq:def_alpha} and \eqref{eq:def_PQ} leads to
\begin{equation}
\label{eq:Sol2lev_LBM}
\begin{cases}
        \displaystyle N_0^{\rm LBM} \!=\! \frac{Q}{(\tau_{\rm esc} \!+\! \Gi \!+\! \Ga) -\Ga\Gs\left(\tau_{\rm esc}^{-1} \!+\! \Gi \!+\! \Gs \!+\! \GEC/2\right)^{-1}}, \\
        \displaystyle N_1^{\rm LBM} = \frac{\Ga N_0^{\rm LB}}{\tau_{\rm esc}^{-1} + \Gi + \Gs + \GEC/2}.
    \end{cases}
\end{equation}
and
\begin{equation}
\label{eq:Sol2lev_DM}
    \begin{cases}
        \displaystyle N_0^{\rm 1D-DM} \!=\! \frac{Q}{\left(\displaystyle \frac{D}{hL} \!+\! \Gi \!+\! \Ga\right) -\frac{\displaystyle \Ga\Gs}{\left(\sqrt{\displaystyle\frac{D\,\GEC}{2h^2}} \displaystyle\coth\!\left(\!\!\sqrt{\frac{L^2\GEC}{2D}}\right) \!+\! \Gi \!+\! \Gs\right)}}, \\[15mm]
        \displaystyle N_1^{\rm 1D-DM} = \frac{\Ga N_0^{\rm DM}}{\sqrt{\displaystyle\frac{D\,\GEC}{2h^2}} \displaystyle\coth\left(\!\sqrt{\frac{L^2\GEC}{2D}}\right) + \Gi + \Gs}.
    \end{cases}
\end{equation}
Note that the latter solution slightly differs from the one given in \cite{2023arXiv230912801B}, where the halving of the decay rate, $\GECtilde=\GEC/2$ in Eq.~\eqref{eq:GEC_halving} for a single electron attached, was not accounted for.

\onecolumn
\setlength{\LTcapwidth}{0.9\textwidth}
\begin{longtable}{lllll}
\caption{List of EC clocks isotopes (i.e., nuclei whose only relevant decay channels is EC) retrieved from \nubase{}~\citep{2021ChPhC..45c0001K}. The columns show the EC clock (formatted as $AXZ$ with $X$ the element name), decay time, stable isotope at the end of its decay chain, and if applicable, the multistep decay chain. A few species with only upper limits on their unmeasured $\beta^+$ channels are also considered as pure EC-unstable: $^{56}{\rm Ni}$ with ${\cal B}r(\beta^+)<\num{5.8e-05}$ ($t_{1/2}> \num{2.9e+04}$\,yr), $^{143}{\rm Pm}$ with ${\cal B}r(\beta^+)<\num{5.7e-06}$ ($t_{1/2}> \num{1.3e+07}$\,yr), and  $^{144}{\rm Pm}$ with ${\cal B}r(\beta^+)<\num{8.0e-05}$ ($t_{1/2}> \num{1.2e+06}$\,yr). \label{tab:EC clocks}}\\[-2.5mm]
\hline\hline
\scriptsize
EC clock &  \multicolumn{2}{c}{$t_{1/2}$ [unit]} & End chain & Multi-step decay path \\
\hline
4Be7   & 53.22  & d   & 3Li7   &  \\
18Ar37 & 35.011 & d   & 17Cl37 &  \\
20Ca41 & 99.4   & kyr & 19K41  &  \\
22Ti44 & 59.1   & yr  & 20Ca44 & $^{44}{\rm Ti}$  [${\rm EC}$, 59.1 yr ] $\to$ $^{44}{\rm Sc}$  [$\beta^{+}$, 4.0421 h ] $\to$ $^{44}{\rm Ca}$ \\
23V49  & 330    & d   & 22Ti49 &  \\
24Cr51 & 27.7015 & d   & 23V51  &  \\
25Mn53 & 3.7    & Myr & 24Cr53 &  \\
26Fe55 & 2.7562 & yr  & 25Mn55 &  \\
28Ni56 & 6.075  & d   & 26Fe56 & $^{56}{\rm Ni}$  [${\rm EC}$, 6.075 d ] $\to$ $^{56}{\rm Co}$  [$\beta^{+}$, 77.236 d ] $\to$ $^{56}{\rm Fe}$ \\
27Co57 & 271.811 & d   & 26Fe57 &  \\
28Ni59 & 81.82  & kyr & 27Co59 &  \\
31Ga67 & 3.2617 & d   & 30Zn67 &  \\
32Ge68 & 271.05 & d   & 30Zn68 & $^{68}{\rm Ge}$  [${\rm EC}$, 271.05 d ] $\to$ $^{68}{\rm Ga}$  [$\beta^{+}$, 67.842 mn ] $\to$ $^{68}{\rm Zn}$ \\
32Ge71 & 11.43  & d   & 31Ga71 &  \\
34Se72 & 8.40   & d   & 32Ge72 & $^{72}{\rm Se}$  [${\rm EC}$, 8.40 d ] $\to$ $^{72}{\rm As}$  [$\beta^{+}$, 26.0 h ] $\to$ $^{72}{\rm Ge}$ \\
33As73 & 80.30  & d   & 32Ge73 &  \\
34Se75 & 119.78 & d   & 33As75 &  \\
36Kr81 & 229    & kyr & 35Br81 &  \\
38Sr82 & 25.35  & d   & 36Kr82 & $^{82}{\rm Sr}$  [${\rm EC}$, 25.35 d ] $\to$ $^{82}{\rm Rb}$  [$\beta^{+}$, 1.2575 mn ] $\to$ $^{82}{\rm Kr}$ \\
37Rb83 & 86.2   & d   & 36Kr83 &  \\
38Sr85 & 64.846 & d   & 37Rb85 &  \\
40Zr88 & 83.4   & d   & 38Sr88 & $^{88}{\rm Zr}$  [${\rm EC}$, 83.4 d ] $\to$ $^{88}{\rm Y}$  [$\beta^{+}$, 106.629 d ] $\to$ $^{88}{\rm Sr}$ \\
42Mo93 & 4.0    & kyr & 41Nb93 &  \\
43Tc97 & 4.21   & Myr & 42Mo97 &  \\
46Pd100 & 3.63   & d   & 44Ru100 & $^{100}{\rm Pd}$  [${\rm EC}$, 3.63 d ] $\to$ $^{100}{\rm Rh}$  [${\rm EC},\beta^{+}$, 1312.30,17.69 mn,d ] $\to$ $^{100}{\rm Ru}$ \\
45Rh101 & 4.07   & yr  & 44Ru101 &  \\
46Pd103 & 16.991 & d   & 45Rh103 &  \\
48Cd109 & 461.3  & d   & 47Ag109 &  \\
50Sn110 & 4.154  & h   & 48Cd110 & $^{110}{\rm Sn}$  [${\rm EC}$, 4.154 h ] $\to$ $^{110}{\rm In}$  [$\beta^{+}$, 4.92 h ] $\to$ $^{110}{\rm Cd}$ \\
49In111 & 2.8048 & d   & 48Cd111 &  \\
52Te118 & 6.00   & d   & 50Sn118 & $^{118}{\rm Te}$  [${\rm EC}$, 6.00 d ] $\to$ $^{118}{\rm Sb}$  [$\beta^{+}$, 3.6 mn ] $\to$ $^{118}{\rm Sn}$ \\
51Sb119 & 38.19  & h   & 50Sn119 &  \\
54Xe122 & 20.1   & h   & 52Te122 & $^{122}{\rm Xe}$  [${\rm EC}$, 20.1 h ] $\to$ $^{122}{\rm I}$  [${\rm EC},\beta^{+}$, 16.50,4.65 mn ] $\to$ $^{122}{\rm Te}$ \\
53I125 & 59.392 & d   & 52Te125 &  \\
54Xe127 & 36.342 & d   & 53I127 &  \\
56Ba128 & 2.43   & d   & 54Xe128 & $^{128}{\rm Ba}$  [${\rm EC}$, 2.43 d ] $\to$ $^{128}{\rm Cs}$  [$\beta^{+}$, 3.640 mn ] $\to$ $^{128}{\rm Xe}$ \\
55Cs131 & 9.689  & d   & 54Xe131 &  \\
56Ba133 & 10.5379 & yr  & 55Cs133 &  \\
58Ce134 & 3.16   & d   & 56Ba134 & $^{134}{\rm Ce}$  [${\rm EC}$, 3.16 d ] $\to$ $^{134}{\rm La}$  [$\beta^{+}$, 6.45 mn ] $\to$ $^{134}{\rm Ba}$ \\
57La137 & 60     & kyr & 56Ba137 &  \\
58Ce139 & 137.642 & d   & 57La139 &  \\
60Nd140 & 3.37   & d   & 58Ce140 & $^{140}{\rm Nd}$  [${\rm EC}$, 3.37 d ] $\to$ $^{140}{\rm Pr}$  [${\rm EC},\beta^{+}$, 6.61,6.96 mn ] $\to$ $^{140}{\rm Ce}$ \\
61Pm143 & 265    & d   & 60Nd143 &  \\
61Pm144 & 363    & d   & 60Nd144 &  \\
61Pm145 & 17.70  & yr  & 60Nd145 &  \\
62Sm145 & 340    & d   & 60Nd145 & $^{145}{\rm Sm}$  [${\rm EC}$, 340 d ] $\to$ $^{145}{\rm Pm}$  [${\rm EC}$, 17.70 yr ] $\to$ $^{145}{\rm Nd}$ \\
64Gd146 & 48.27  & d   & 60Nd142 & $^{146}{\rm Gd}$  [${\rm EC}$, 48.27 d ] $\to$ $^{146}{\rm Eu}$  [$\beta^{+}$, 4.61 d ] $\to$ $^{146}{\rm Sm}$  [$\alpha$, 68 Myr ] $\to$ $^{142}{\rm Nd}$ \\
63Eu149 & 93.1   & d   & 62Sm149 &  \\
64Gd153 & 240.6  & d   & 63Eu153 &  \\
65Tb155 & 5.32   & d   & 64Gd155 &  \\
65Tb157 & 71     & yr  & 64Gd157 &  \\
68Er158 & 2.29   & h   & 66Dy158 & $^{158}{\rm Er}$  [${\rm EC}$, 2.29 h ] $\to$ $^{158}{\rm Ho}$  [$\beta^{+}$, 11.3 mn ] $\to$ $^{158}{\rm Dy}$ \\
66Dy159 & 144.4  & d   & 65Tb159 &  \\
68Er160 & 28.58  & h   & 66Dy160 & $^{160}{\rm Er}$  [${\rm EC}$, 28.58 h ] $\to$ $^{160}{\rm Ho}$  [$\beta^{+}$, 25.6 mn ] $\to$ $^{160}{\rm Dy}$ \\
67Ho161 & 2.48   & h   & 66Dy161 &  \\
67Ho163 & 4.570  & kyr & 66Dy163 &  \\
70Yb164 & 75.8   & mn  & 68Er164 & $^{164}{\rm Yb}$  [${\rm EC}$, 75.8 mn ] $\to$ $^{164}{\rm Tm}$  [${\rm EC},\beta^{+}$, 3.28,5.13 mn ] $\to$ $^{164}{\rm Er}$ \\
68Er165 & 10.36  & h   & 67Ho165 &  \\
70Yb166 & 56.7   & h   & 68Er166 & $^{166}{\rm Yb}$  [${\rm EC}$, 56.7 h ] $\to$ $^{166}{\rm Tm}$  [$\beta^{+}$, 7.70 h ] $\to$ $^{166}{\rm Er}$ \\
69Tm167 & 9.25   & d   & 68Er167 &  \\
70Yb169 & 32.014 & d   & 69Tm169 &  \\
72Hf170 & 16.01  & h   & 70Yb170 & $^{170}{\rm Hf}$  [${\rm EC}$, 16.01 h ] $\to$ $^{170}{\rm Lu}$  [$\beta^{+}$, 2.012 d ] $\to$ $^{170}{\rm Yb}$ \\
72Hf172 & 1.87   & yr  & 70Yb172 & $^{172}{\rm Hf}$  [${\rm EC}$, 1.87 yr ] $\to$ $^{172}{\rm Lu}$  [$\beta^{+}$, 6.70 d ] $\to$ $^{172}{\rm Yb}$ \\
71Lu173 & 1.37   & yr  & 70Yb173 &  \\
72Hf175 & 70.65  & d   & 71Lu175 &  \\
74W176 & 2.5    & h   & 72Hf176 & $^{176}{\rm W}$  [${\rm EC}$, 2.5 h ] $\to$ $^{176}{\rm Ta}$  [$\beta^{+}$, 8.09 h ] $\to$ $^{176}{\rm Hf}$ \\
74W178 & 21.6   & d   & 72Hf178 & $^{178}{\rm W}$  [${\rm EC}$, 21.6 d ] $\to$ $^{178}{\rm Ta}$  [$\beta^{+}$, 2.36 h ] $\to$ $^{178}{\rm Hf}$ \\
73Ta179 & 1.82   & yr  & 72Hf179 &  \\
74W181 & 120.956 & d   & 73Ta181 &  \\
76Os182 & 21.84  & h   & 74W182 & $^{182}{\rm Os}$  [${\rm EC}$, 21.84 h ] $\to$ $^{182}{\rm Re}$  [$\beta^{+}$, 64.2 h ] $\to$ $^{182}{\rm W}$ \\
75Re183 & 70.0   & d   & 74W183 &  \\
76Os185 & 92.95  & d   & 75Re185 &  \\
77Ir189 & 13.2   & d   & 76Os189 &  \\
78Pt191 & 2.83   & d   & 77Ir191 &  \\
80Hg192 & 4.85   & h   & 78Pt192 & $^{192}{\rm Hg}$  [${\rm EC}$, 4.85 h ] $\to$ $^{192}{\rm Au}$  [$\beta^{+}$, 4.94 h ] $\to$ $^{192}{\rm Pt}$ \\
78Pt193 & 50     & yr  & 77Ir193 &  \\
80Hg194 & 447    & yr  & 78Pt194 & $^{194}{\rm Hg}$  [${\rm EC}$, 447 yr ] $\to$ $^{194}{\rm Au}$  [$\beta^{+}$, 38.02 h ] $\to$ $^{194}{\rm Pt}$ \\
79Au195 & 186.01 & d   & 78Pt195 &  \\
80Hg197 & 64.93  & h   & 79Au197 &  \\
82Pb200 & 21.5   & h   & 80Hg200 & $^{200}{\rm Pb}$  [${\rm EC}$, 21.5 h ] $\to$ $^{200}{\rm Tl}$  [$\beta^{+}$, 26.1 h ] $\to$ $^{200}{\rm Hg}$ \\
81Tl201 & 3.0421 & d   & 80Hg201 &  \\
81Tl202 & 12.31  & d   & 80Hg202 &  \\
82Pb202 & 52.5   & kyr & 80Hg202 & $^{202}{\rm Pb}$  [${\rm EC}$, 52.5 kyr ] $\to$ $^{202}{\rm Tl}$  [${\rm EC}$, 12.31 d ] $\to$ $^{202}{\rm Hg}$ \\
82Pb203 & 51.924 & h   & 81Tl203 &  \\
82Pb205 & 17.0   & Myr & 81Tl205 &  \\
\hline
\end{longtable}

\twocolumn


\end{appendix}

\bibliography{ec_revisited}

\end{document}